\documentclass[prx,aps,floatfix,longbibliography,superscriptaddress,twocolumn,final,notitlepage]{revtex4-2}
\usepackage[utf8]{inputenc}
\usepackage[T1]{fontenc}
\usepackage{multirow}
\usepackage{subcaption}
\usepackage{tabularx} 
\usepackage{graphicx,caption}
\usepackage{amsfonts,amsmath,amssymb}
\usepackage{comment}
\usepackage{lineno}
\usepackage{hyperref}
\usepackage[explicit]{titlesec}
\usepackage[nameinlink]{cleveref}
\Crefname{figure}{Fig.}{Figs.}% {<type>}{<singular>}{<plural>}
\Crefname{equation}{Eq.}{Eqs.}% {<type>}{<singular>}{<plural>}

\newcommand{\Exp}{\mathbb{E}}
\newcommand{\pois}{\textrm{Pois}}

\newcommand{\be}{\begin{equation}}
\newcommand{\ee}{\end{equation}}
\newcommand{\bea}{\begin{eqnarray}}
\newcommand{\eea}{\end{eqnarray}}

\newcommand{\f}[2]{\frac{#1}{#2}}
\newcommand{\bup}[1]{\left(#1\right)}
\newcommand{\rup}[1]{\left[#1\right]}

\renewcommand{\ref}[1]{[\ref{#1}]}
\usepackage{tabularx}

\usepackage{amsfonts,amsmath,amssymb}   % clashes with mathdesign
\usepackage{comment}
\usepackage{lineno}
\usepackage{hyperref}
\usepackage[explicit]{titlesec}
\usepackage{graphicx}
\usepackage[labelfont=bf]{caption}
\usepackage[format=hang]{subcaption}
\usepackage{algorithm,algorithmic}
\usepackage{amsmath,fixmath}

\usepackage[algoruled,algo2e]{algorithm2e}
\usepackage{sidecap}
\usepackage{boldline}
\usepackage{setspace}
\usepackage{listings}
\usepackage{fancyvrb}
\usepackage{soul}

\usepackage{mathtools}

\newcommand{\ULsubfloat}[2][\empty]% #1 = caption (optional), #2 = image
{\hbox{% unbreakable group
  \sbox0{#2}% measure height of image
  \captionsetup{position=top, justification=centering, singlelinecheck=false}%
  \rotatebox[origin=bl]{90}{\begin{minipage}[b]{\dimexpr \ht0+\dp0}
    \subcaption*{#1}
  \end{minipage}}\raisebox{\dp0}{\usebox0}%
}}

\usepackage[table]{}
\usepackage{array}

\usepackage[utf8]{inputenc}

\usepackage{booktabs}
\usepackage{adjustbox}
\usepackage{changepage}

\usepackage[usenames,dvipsnames]{xcolor}
\definecolor{shadecolor}{gray}{0.9}
\usepackage{tabu}

\usepackage[normalem]{ulem} % to use nix

\newcommand{\mtrep}{\mbox{{\small CRep}}}
\newcommand{\mt}{\mbox{{\small MT}}}
\newcommand{\bpmf}{\mbox{{\small BPMF}}}
\newcommand{\mtrepo}{\mbox{{\small CRep$_0$}}}
\newcommand{\mtrepnc}{\mbox{{\small CRep$_{nc}$}}}
\newcommand{\leiden}{\mbox{{\small Leiden}}}
\newcommand{\olp}{\mbox{{\small OLP}}}
\graphicspath{{./figures/}} % Directory in which figures are stored
\usepackage{enumitem,amssymb}
\newlist{todolist}{itemize}{2}
\setlist[todolist]{label=$\square$}
\newlist{todolist_done}{itemize}{2}
\setlist[todolist_done]{label=$\blacksquare$}
\usepackage{pifont}
\begin{document}

\title{Generative model for reciprocity and community detection in networks}

\author{Hadiseh Safdari}
	\email{hadiseh.safdari@tuebingen.mpg.de}
%	\thanks{Contributed equally.}
	\affiliation{Max Planck Institute for Intelligent Systems, Cyber Valley, Tuebingen 72076, Germany}

\author{Martina Contisciani}
	\email{martina.contisciani@tuebingen.mpg.de}
	\thanks{Contributed equally.}
	\affiliation{Max Planck Institute for Intelligent Systems, Cyber Valley, Tuebingen 72076, Germany}

\author{Caterina De Bacco}
	\email{caterina.debacco@tuebingen.mpg.de}
%	\thanks{Contributed equally.}
	\affiliation{Max Planck Institute for Intelligent Systems, Cyber Valley, Tuebingen 72076, Germany}

% Optional adjustment to line up main text (after abstract) of first page with line numbers, when using both lineno and twocolumn options.
% You should only change this length when you've finalised the article contents.
%\verticaladjustment{-2pt}

%\thispagestyle{firststyle}
%\ifthenelse{\boolean{shortarticle}}{\ifthenelse{\boolean{singlecolumn}}{\abscontentformatted}{\abscontent}}{}

%%%%%%%%%%%%%%%%%%%%%%%%%%%%%%%%%%%%%%%%%%%%%%%%%%%%%%%%%%
\begin{abstract} % MAX 150 words
We present a probabilistic generative model and efficient algorithm to model reciprocity in directed networks. Unlike other methods that address this problem such as exponential random graphs, it assigns latent variables as community memberships to nodes and a reciprocity parameter to the whole network rather than fitting order statistics. It formalizes the assumption that a directed interaction is more likely to occur if an individual has already observed an interaction towards her. It provides a natural framework for relaxing the common assumption in network generative models of conditional independence between edges, and it can be used to perform inference tasks such as predicting the existence of an edge given the observation of an edge in the reverse direction. Inference is performed using an efficient expectation-maximization algorithm that exploits the sparsity of the network, leading to an efficient and scalable implementation. We illustrate these findings by analyzing synthetic and real data, including social networks, academic citations and the Erasmus student exchange program. Our method outperforms others in both predicting edges and generating networks that reflect the reciprocity values observed in real data, while at the same time inferring an underlying community structure. We provide an open-source implementation of the code online.
\end{abstract}

\maketitle

\section{Introduction}
 Reciprocity in directed networks, i.e., the tendency of a pair of nodes to form mutual connections between each other
 \cite{wasserman1994social}, is an important feature of many social relationships. Its impact ranges from affecting the development of exchange and power to determining the emergence of trust and solidarity \cite{molm2010structure,nowak2005evolution}. Behavior of this kind has also been found in many kinds of networks that reflect human and institutional interaction, e.g.,  the  world  wide web, online dating, interfirm contracts, journal citations and  email communication \cite{garlaschelli2005,zhao2013user,wincent2010quality,li2019reciprocity,newman2002email}. 

Among the various network modeling approaches, that of probabilistic generative models enable us for a rigorous theoretical foundation within the framework of statistical inference, as well as a flexible incorporation of domain knowledge in the modeling assumptions. Here, we consider a latent variable model, a probabilistic approach that contains latent and observed variables. The latent variables encode hidden patterns in the data, such as community memberships, and determine the probability of ties between nodes. For instance, knowing which communities two nodes belong to helps determine the likelihood of their interaction. 

While in some simple cases, community structure may explain the tendency toward reciprocation \cite{holland1983stochastic}, this mechanism may not be enough to capture more complex scenarios. Indeed, many generative models with community structure fail to reproduce the values of reciprocity observed in real networks, as we discuss in more details later. Conversely, several models aimed at capturing reciprocity do not account for community structure \cite{holland1981exponential,park2004statistical}. It is reasonable to expect that the mechanism regulating the existence of interactions can be influenced by both patterns of communities and reciprocity. In addition, communities are often interpretable objects and may correspond to functional unit, hence the value of including them in the model formulation.
Incorporating reciprocity as well as community structure into a coherent latent variable model comes with the main challenge of relaxing the conditional independence assumption between edges, a common assumption in generative models to ease mathematical derivations. In addition, this task requires properly capturing conditional probabilities, as we describe later.
Inspired by these insights, we propose a novel probabilistic latent variable approach to model networks that preserves the benefits of generative models, while capturing both community structure and reciprocity.

% -------- Related work -------
%\subsection*{Related work}
Models for reciprocity and latent community structure have largely been developed independently of one another, and only a handful of works have hinted at incorporating them into a unique framework. 
For instance, Garlaschelli and Loffredo \cite{garlaschelli2006} point towards a possible relationship between their model for reciprocity and general hidden variable models.
%For instance, Holland et al.~\cite{holland1983stochastic} describe a stochastic block model with reciprocity, while Garlaschelli and Loffredo \cite{garlaschelli2006} point towards a possible relationship between their model for reciprocity and general hidden variable models. 
Most notably, the pair-dependent stochastic block model of Holland et al. \cite{holland1983stochastic}, well explained also by Wasserman and Anderson \cite{wasserman1987stochastic}, holds assumptions similar to ours, in that it models  jointly pairs of edges, which they call dyad vectors. While a seminal work, it is, nevertheless, limited to   hard membership and unweighted networks; hence the likelihood function that they propose substantially differs from the likelihood represented by our model. One practical aspect of our choice for the likelihood is that parameters' inference in our model is optimized to fully exploit the sparsity of the dataset and is scalable to large network sizes.   

Reciprocity is often modeled by means of exponential random graphs \cite{holland1981exponential,park2004statistical,robins2007introduction,squartini2013reciprocity}, where it is treated as a measured network property that needs to be reproduced (often together with other network properties like the degree) by sampling networks using statistical mechanics principles, e.g.,  maximum entropy.
The approach presented in this work significantly differs from  the previous studies in that we include latent variables, such as community membership, as a mechanism to determine edge formation.   However,  in the case of  exponential random graphs,  possible group structures are not given a priori as the latent parameters; instead, they can only be estimated a posteriori on the sampled networks. More broadly, our approach is that of generative models, which incorporate a priori community structure by means of latent variables, and these are inferred from the observed interactions \cite{de2017community,ball2011efficient}. However, in these generative models reciprocity is not explicitly included as a mechanism for tie formation, thus these models often fail to reproduce the observed reciprocity values of real networks.
Consequently,  a generative method whose latent variables describe both reciprocity and community memberships is needed.

 %This approach does not include latent variables such as community membership; therefore possible group structures can only be estimated a posteriori on the sampled networks. In contrast, generative models incorporate a priori community structure by means of latent variables, which are inferred from the observed interactions \cite{de2017community,ball2011efficient}.

\section{Relaxing the conditional independence assumption}
A possible explanation for the practical deficiency of generative models with communities to reproduce observed reciprocity values is the common assumption of conditional independence between edges, which makes the problem both analytically and computationally more tractable. This assumption states that the likelihood of a directed tie between two nodes depends only on their community membership (and other possible model parameters), but not on the existence of the reciprocated edge. This might be too strict of an assumption to capture the feature of reciprocity, where it is reasonable to expect that the existence of an edge in one direction should also be conditioned  on the existence of an edge in the opposite direction. For instance, if an author $i$ has  cited another author $j$, this might predict the probability of  $j$ also citing $i$.
At the same time, knowing the communities that the authors belong to, could also help estimating this probability.  Mathematically, this can be translated to relaxing the assumption of conditional independence, which is the approach we take here.

Formally, we represent interactions between $N$ individuals as a weighted asymmetric matrix $A$, with entries $A_{ij}$ being the number (or weight) of interactions from $i$ to $j$; for instance, the number of favors or services that $i$ does for $j$, or the number of times that $i$ has endorsed $j$, e.g., as paper citations.
Our model assigns a \textit{joint} likelihood $P(A_{ij},A_{ji}|\Theta)$ to edges involving the same pairs of nodes $(i,j)$, given some set of latent parameters $\Theta$. Specifically, we assume the likelihood of a network to factorize as:
  \be\label{eqn:likL}
 P(A|\Theta) = \prod_{i<j} P(A_{ij},A_{ji}|\Theta) \quad.
\ee
This is fundamentally different from the prevalent approaches in generative models, where, typically, one assumes that \textit{individual} edges are conditionally independent given the network parameters, i.e.,  $P(A|\Theta) = \prod_{i,j} P(A_{ij}|\Theta)$.\\

Notice that edges involving different pairs of nodes remain conditionally independent as in standard approaches. Equivalently, in terms of the conditional distribution of an individual edge $P(A_{ij}| A_{ji},\Theta)$, we assume that this can be different than its marginal distribution $P(A_{ij}|\Theta)$.
To the extent of our knowledge, this assumption has never been deeply questioned, except for a few works \cite{lloyd2012random,orbanz2014bayesian}. As firstly pointed out by Hoff \cite{hoff2008modeling}, there are theoretical groundings for this assumption to hold in common scenarios, due to generalizations of de Finetti's theorem by Aldous \cite{aldous1981representations} and Hoover \cite{hoover1979relations} (see \cite{orbanz2014bayesian} for a detailed discussion). They show that, for exchangeable graphs, i.e.,  in networks without any natural order between nodes (which is often the case), the joint probability function of the adjacency entries can be properly represented using latent variables on nodes and pairs. In other words,  the joint can be factorized as a product on edges, given the latent variables. 

However, in the case of directed networks, where the adjacency matrix is  asymmetric, as in our case, a precise representation can only be obtained using \Cref{eqn:likL}. While standard conditionally independent models can in principle arbitrarily well approximate the whole network distribution \cite{kallenberg1999multivariate}, in practice it is not known how state-of-the-art models perform on this regard.
To effectively model reciprocity, we relax the assumption of conditional independence and include the pairwise dependencies of two directed edges between pairs of nodes; such minimal relaxation is required to effectively model reciprocity. We compare results against standard conditionally independent models in terms of various performance metrics on both synthetic and real data.

%%%%%%%%%%   The CRep model  %%%%%%%%%% %%%%%%%%%% %%%%%%%%%% %%%%%%%%%%

\section{The community-reciprocity model}
\label{sec:reciprocity}
To fully specify the joint likelihood in Eq.~(\ref{eqn:likL}), we need to characterize conditional distributions and one-point marginals like the distribution $P(A_{ij}| A_{ji},\Theta)$ and $P(A_{ij}|\Theta)$.  Here, we aim at capturing reciprocity, hence we assume that observed interactions exist because of two types of contributions:  i) the communities that nodes belong to, as in general community detection frameworks like the stochastic block model \cite{holland1983stochastic}, and ii) the fact that an individual that receives a directed interaction is more likely to reciprocate.
In order to construct a model  flexible enough to capture weighted networks and overlapping communities, we utilize  a mixed-membership approach, similar to \cite{ball2011efficient,de2017community}, to model how communities regulate edge formation.

Given the adjacency matrix $A$, our goal is to find community memberships of nodes and the magnitude of the reciprocity effect in the network, i.e., $\Theta$.
Bringing the contributions of reciprocity and community structure together, we model the conditional probability of $A_{ij}$ given $A_{ji}$ as drawn from a Poisson distribution
\bea
P(A_{ij}| A_{ji},\Theta)&=&   \frac{e^{-\lambda_{ij}} \, \lambda_{ij}^{A_{ij}}}{A_{ij}!}\quad,
 \label{eq:poiss_dist}
\eea
with mean
\be \label{eq:meanPoissonC}
\lambda_{ij}=\lambda_{ij}^{0}\,+\,  \eta A_{ji}=\bup{\sum_{k,q=1}^{K} u_{ik}v_{jq}w_{kq}} + \eta \, A_{ji} \quad.
\ee
We denote with $\Theta=(u,v,w,\eta)$  the latent parameters that we want to infer. The parameters $u_{ik}$, $v_{ik}$ are entries of $K$-dimensional vectors $u_{i}$ and $v_{i}$, the out-going and in-coming communities respectively; $w_{kq}$ are the entries of a $K\times K$ affinity matrix, which regulates the structure of communities, e.g., assortative when its diagonal entries are greater than off-diagonal entries, in this case edges are more likely between nodes in the same community; $\eta$ is the reciprocity parameter, and it regulates the  impact of observing $A_{ji}$ to predict the existence of $A_{ij}$. We omit from it the number of communities $K$, as in this work we assume this as given. When unknown, as in our experiments with real data, we estimate it by using cross-validation.\\

Notice that $\lambda_{ij}$ includes separate contributions from both community parameters and reciprocity coefficient. It assumes additive contributions: we can have zero contribution from one term and still observe the existence of an edge because of the other term. If both are non-zero, their total impact sums up. This is conceptually different than a multiplicative contribution,   a possible modeling choice that we do not explore here. 
Intuitively, an edge with weight $A_{ij}$ exists if $i$ and $j$ belong to compatible communities (compatibility is regulated by the affinity matrix) or because of the reciprocity effect of observing the opposite edge $A_{ji}$. For instance, an author might cite another one because they belong to the same community (e.g., a research sub-field) or because she was cited by the other on a previous publication.

Finally, as we need positive $\lambda_{ij}$, we assume $\eta \geq 0$. This restricts the model to have positive reciprocity contribution, i.e., receiving an in-coming edge can only boost the likelihood of the corresponding out-going edge, but not decrease it. Although this assumption could be limiting in certain contexts, it nevertheless applies to several relevant scenarios, in particular to the cases we study here. Relaxing this assumption, and suitably modifying the underlying theoretical model,  is left for future works.

\begin{figure}[h]
	\includegraphics[width=1\linewidth]{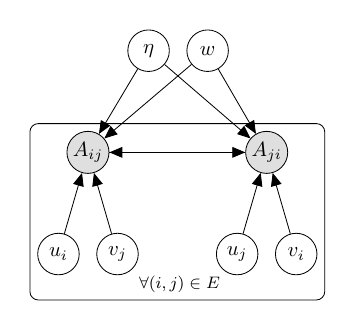}
	\caption{
	\textbf{Graphical model representation.}  $A_{ij}$ and $A_{ji}$ are the edges involving the same pairs of nodes $(i, j)$; $\eta, w, u, v$ are the latent parameters $\Theta$; $E$ denotes the set of network edges.}
	\label{fig:graphicalmodel}
\end{figure}

Our model specifies conditional probabilities,  however, we do not assume the existence of a consistent joint distribution. In fact, finding a closed-form for the joint in Eq.~(\ref{eqn:likL}), consistent with our proposed conditional, requires specifying a marginal probability function and then enforce consistency equations like $\sum_{A_{ji}} P(A_{ij}|A_{ji},\Theta)\,P(A_{ji}|\Theta)=P(A_{ij}|\Theta)$. Depending on the choice of this marginal, enforcing consistency might be non-trivial, as it may require performing intractable marginalization.  
 Early formalizations of the consistency between conditional and joint distribution has been provided, in a seminal work,  by Besag’s Auto-Poisson models \cite{besag1974spatial}. In the context of graphical models, a few models specify conditional Poisson distributions \cite{allen2013local,hadiji2015poisson}, but without considering latent variables.
In the absence of a closed-form joint distribution, we adopt a tractable pseudo-likelihood approach \cite{besag1974spatial}, where instead of optimizing the exact likelihood of Eq.~(\ref{eqn:likL}), we consider the approximation:
\be\label{eqn:PSL}
P(A|\Theta)= \prod_{i<j} P(A_{ij},A_{ji}|\Theta) \approx \prod_{i,j}P(A_{ij}|A_{ji},\Theta) \quad,%\, P(A_{ji}|A_{ij},\Theta)
\ee
 which is available in closed-form as it requires only the conditional probabilities, which we specified above. The equality holds only when $A_{ij}$ and $A_{ji}$ are conditionally independent, the common assumption in network generative models, as in that case $P(A_{ij}|A_{ji},\Theta)=P(A_{ij}|\Theta)$. This is not our case since we relax this assumption, and Eq.~(\ref{eqn:PSL}) is only an approximation. This approach has also been considered in dyadic-dependent models \cite{strauss1990pseudolikelihood}, for community detection in networks \cite{amini2013pseudo}, and for local Poisson graphical models \cite{allen2013local}. A visual overview of our model is shown in \Cref{fig:graphicalmodel}.

%%   --------------------------------------------------------------------------------------------------------------------------------------------------------------------------

%   --------------------------------------------------------------------------------------------------------------------------------------------------------------------------

 % ------- ------- ------- 				GENERATIVE MODEL 		------- ------- ------- -------
\section{Inference with expectation-maximization}
\label{sec:em}
The goal is to find the community and reciprocity parameters, i.e., $\Theta$, given the adjacency matrix.
Defining $L^{ps}_{ij}(\Theta,A_{ji})=\log  P(A_{ij}|A_{ji}, \Theta)$ and neglecting the factorial term which is  independent of these parameters, we have the log-pseudo-likelihood:
\bea\label{eqn:LogPSL1}
L^{ps}(\Theta)&=&\sum_{i,j}L^{ps}_{ij}(\Theta)=\sum_{i,j}\bup{ A_{ij}  \log\lambda_{ij} -\lambda_{ij}} \quad.
\eea
We aim at maximizing this quantity, but the presence of the logarithmic term makes this maximization difficult. However, using a variational approach by means of Jensen's inequality, it can be shown (see \Cref{apxsubsec:VI}) that  maximizing $L^{ps}(\Theta)$ is equivalent to maximizing
\bea\label{eqn:LogPSL2}  
L^{ps}(\Theta, \rho,\phi) &=& \sum_{i,j} \left\{ A_{ij}    \,
 \rho^{(1)}_{ij}\left( \sum_{k,q}  \phi _{ijkq} \log { u_{ik} v_{jq} w_{kq} }\, \right. \right.  \nonumber \\
 && \left. \left. -\sum_{k,q}  \phi _{ijkq}\log{ \phi _{ijkq}} \right) \,+ A_{ij}  \,  \rho^{(2)}_{ij }  \log { \eta A_{ji}}   \right. \nonumber  \\
&&-\,A_{ij}  \,  \left(\rho^{(1)}_{ij} \log{\rho^{(1)}_{ij }}+\rho^{(2)}_{ij } \log{\rho^{(2)}_{ij }} \right)\nonumber \\
&& \left. -\sum_{k,q}u_{ik}v_{jq}w_{kq}  -\,  \eta A_{ji} \right \} \quad,
\eea
with respect to $\Theta$, $\rho=\bup{\rho^{(1)}, \rho^{(2)}}$, and  $\phi$, where
\be   \label{eqn:rho}
  \rho^{(1)}_{ij}= \frac{\lambda^{0}_{ij}}{\lambda^{0}_{ij}+\eta\,A_{ji}}  \;,\quad \rho^{(2)}_{ij}= \frac{\eta\,A_{ji}}{\lambda^{0}_{ij}+\eta\,A_{ji}}  \,,
\ee

\be \label{eqn:phi}
        \phi_{ijkq} =\frac{u_{ik}v_{jq}w_{kq} }{\lambda^{0}_{ij}} \quad ,
\ee
%\bea
%  \rho^{(1)}_{ij}&=& \frac{\lambda^{0}_{ij}}{\lambda^{0}_{ij}+\eta\,A_{ji}} \label{eqn:rho1}\\
%    \rho^{(2)}_{ij}&=& \frac{\eta\,A_{ji}}{\lambda^{0}_{ij}+\eta\,A_{ji}} \label{eqn:rho2}\\
%        \phi_{ijkq} &=&\frac{u_{ik}v_{jq}w_{kq} }{\lambda^{0}_{ij}} \label{eqn:phi} \quad.
%\eea
are the variational distributions over the parameters.

Constraints on the parameters can be arbitrarily added, e.g., $\sum_{k}u_{ik}=\sum_{k}v_{ik}=1$, by incorporating Lagrange multipliers inside Eq. (\ref{eqn:LogPSL1}), and repeating similar calculations. In our numerical experiments,  we consider both constrained and unconstrained cases.

 We can perform this optimization by alternatively updating the various parameters, with an expectation-maximization (EM) algorithm.
At each step, one updates $\rho$ and $ \phi$ using Eqs.~(\ref{eqn:rho})-(\ref{eqn:phi}) (E-step) and then maximizes $L^{ps}(\Theta, \rho,\phi)$ with respect to $\Theta$ by setting partial derivatives to zero (M-step). This iteration is repeated until $L^{ps}$ convergences.
The whole routine is described in \Cref{alg:EM} and the detailed derivations are in the \Cref{apx:derivations}. This algorithm is computationally efficient and scalable to large system sizes  as it exploits the sparsity of the dataset. Indeed, all the updates involve in the numerator sums over $A_{ij}$, hence only the non-zero entries count, giving an algorithmic complexity of  $O(M\, K^{2})$.% for a total of $2\,N\, K+K^{2}+1$ parameters.
%\bea
%\eta^{(t+1)}&=&\f{ \sum_{ij} A_{ij} \rho^{(2)}_{ij}}{\sum_{ij} A_{ji}} = \f{\sum_{ij}A_{ij} \bup{\frac{\eta^{(t)}\,A_{ji}}{\lambda^{0,(t)}_{ij}+\eta^{(t)}\,A_{ji}} }}{M}\\
%&=& \f{ \eta^{(t)} }{M}\,\sum_{ij}\f{A_{ij}A_{ji} }{\lambda^{(t)}_{ij}}\quad.
%\eea

%% ------------------------------ Algorithm ------------------------------------------------------------
\setlength{\textfloatsep}{5pt}
\begin{algorithm}[H]
\SetKwInOut{Input}{Input}
	\setstretch{0.7}
	\Input{network $A=\{A_{ij}\}_{i,j=1}^{N}$, \\number of communities $K$.}
  	\BlankLine
	\KwOut{membership vectors $u=\rup{u_{ik}},\, v=\rup{v_{ik}}$; network-affinity matrix $w=\rup{w_{kq}}$; reciprocity parameter $\eta$.}
	\BlankLine
	 Initialize $u,v,w,\eta$ at random. 
	 \BlankLine
	 Repeat until $L^{ps}$ convergences:
	 \BlankLine
	\quad 1. Calculate $\rho^{(1)}$ and $\phi$ (E-step): %using Eq. (\ref{eqn:Estep}) 
	\be
	  \rho^{(1)}_{ij}= \frac{\lambda^{0}_{ij}}{\lambda^{0}_{ij}+\eta\,A_{ji}}\;, \quad \phi_{ijkq} =\frac{u_{ik}v_{jq}w_{kq}}{\lambda_{ij}^{0}}%\rho^{(2)}_{ij}= \frac{\eta\,A_{ji}}{\lambda^{0}_{ij}+\eta\,A_{ji}} \;,
	\nonumber
	\ee
%	\be
%	\phi_{ijkq} =\frac{u_{ik}v_{jq}w_{kq}}{\lambda_{ij}^{0}} \nonumber \quad.
%	\ee
	 \quad 2. Update parameters $\Theta$ (M-step):  
	\BlankLine
	\quad \quad \quad 
		i) for each node $i$ and community $k$ update memberships:
		\bea
		\quad  u_{ik}&=& \frac{1}{\gamma^{u}_i} \sum_{j,q} A_{ij} \rho^{(1)}_{ij}\phi_{ijkq}  \nonumber\\
\quad  v_{ik}&=& \frac{1}{\gamma^{v}_i} \sum_{j,q} A_{ji} \rho^{(1)}_{ji}\phi_{jiqk} \nonumber
		\eea
	\quad \quad \quad
	ii) for each pair $(k,q)$ update affinity matrix:
		\be
		\quad w_{kq} =\frac{\sum_{i,j } A_{ij} \rho^{(1)}_{ij}\phi_{ijkq} }{\sum_{i,j} u_{ik}\, v_{jq}} \nonumber
		 \ee
	\quad \quad \quad
%		ii) if $\eta>0$ update reciprocity parameter:
		iii) update reciprocity parameter:
		\be
		\quad \eta = \f{ \eta}{M}\,\sum_{i,j}\f{A_{ij}A_{ji} }{\lambda_{ij}} \nonumber
		 \ee
Note: $\gamma_{i}^{u},\gamma_{i}^{v}$ are quantities that are defined differently for constrained and unconstrained values of $u_{i}$ and $v_{i}$. In the constrained case, they correspond to Lagrange multipliers; see \Cref{apxsubsec:EM}.
	\caption{\mtrep:  EM algorithm}
	\label{alg:EM}
\end{algorithm}
%% ------------------------------------------------------------------------------------------

%%%%%%%%%% -----------------------------------------------------------------
\section{A benchmark generative model with communities and reciprocity}
So far we have focused on recovering the  model parameters  given the data, i.e., the inference.
In this section, instead, we propose a benchmark probabilistic generative model to generate synthetic data with intrinsic community structure, and a given reciprocity value.  It takes as input a set of membership vectors, $u_{i}$ and $v_{i}$,  affinity matrix $w$, and reciprocity parameter $\eta$; the output is a directed network with adjacency matrix $A$. In this formulation, edges between a given pair of nodes are  generated stochastically; one edge being generated first and independent from the other, while the formation of the opposite edge depends on how the first was drawn. The pairs of edges are conditionally independent from each other.
Formally, we aim at sampling pairs of edges from Eq. (\ref{eqn:likL}), which can be done by selecting a marginal $P(A_{ij}|\Theta)$ and a conditional distribution $P(A_{ji}| A_{ij}, \Theta)$.
By assuming a Poisson conditional as in  Eq. (\ref{eq:poiss_dist}) and a Poisson marginal, our model would reduce  to a standard (conditionally independent) generative model with communities in the case of zero reciprocity parameter.  Even though with this choice the joint is computationally intractable, this is not an issue, as we do not aim to use the joint to compute quantities analytically, but rather focus on sampling from it.
Formally, given the input set of latent variables $\Theta=(u,v,w,\eta)$, we draw a pair $(A_{ij},A_{ji})$ consistently with the joint $P(A_{ij},A_{ji}|\Theta)$, in a two-step sampling routine:
\begin{enumerate}
\item Select with a coin-flip one direction, $(i,j)$ or $(j,i)$. Say we select $(i,j)$.
\item Sample $A_{ij}$ from the marginal %P(A_{ij}|\Theta)%
\bea
P(A_{ij}|\Theta)= \pois(m_{ij}) \quad, 
\eea
where
\bea\label{eqn:mij}
m_{ij}=\f{ \lambda ^{0}_{ij} +\eta \lambda ^{0}_{ji}}{ \bup{1- \eta^{2}}}
\eea
is the mean of the marginal distribution such that it is consistent with the joint and the conditional distributions. Indeed,  $\Exp \rup{A_{ij}} = m_{ij} = \sum_{A_{ij}} A_{ij} \, P(A_{ij}|\Theta) = \sum_{A_{ij},A_{ji}} A_{ij} \, P(A_{ij},A_{ji}|\Theta)$ (see \Cref{apxsubsec:meanMarginal} for the complete derivation).
\item Sample $A_{ji}$ from the conditional %P(A_{ji}|A_{ij}, \Theta)%
\bea
P(A_{ji}|A_{ij}, \Theta)= \pois(\lambda_{ji}^{0}\,+\,  \eta\, A_{ij}) \quad,
\eea using the previously extracted value of $A_{ij}$.
\end{enumerate}

The Poisson distribution may generate multiple edges between a pair of nodes, so this model may create multigraphs.  This is consistent with the interpretation that $A_{ij}$ is the number, or total weight, of links from $i$ to $j$.  If we wish to generate binary networks where $A_{ij} \in \{0,1\}$, we use the fact that the Poisson and Bernoulli distributions become close in the sparse limit.
To enforce sparsity,  it is sufficient to multiply the $\lambda^{0}_{ij}$ by a constant $\zeta$, as the $m_{ij}$ in Eq. (\ref{eqn:mij}) will also be automatically rescaled by the same quantity. The constant can be fixed by choosing a value for the expected number of (weighted) edges:
\bea
\Exp\rup{M} &=& \sum_{i,j}\f{ \zeta\, \lambda_{ij}^{0}+\zeta\,\eta\, \lambda_{ji}^{0}}{1-\eta^{2}}=\f{\zeta}{1-\eta}\, \sum_{i,j}\lambda^{0}_{ij} \\
\rightarrow \zeta &=& (1-\eta)\, \f{\Exp\rup{M} }{\sum_{i,j}\lambda^{0}_{ij}} \quad.
\eea

Imagine now a practitioner willing to control for the relative contribution of community and reciprocity in generating edges. Our model naturally allows this possibility, as this tuning  is encoded by $\eta$. To see this explicitly, we calculate the fraction of edges generated by community effects only and introduce the $cr_{ratio}$ variable as following:
\be\label{eqn:crratio}
cr_{ratio} := \f{\sum_{i,j}\lambda^{0}_{ij}}{\Exp\rup{M}} = 1-\eta \quad,
\ee
where we used Eq.~(\ref{eqn:mij}) to rewrite the denominator. Thus, by varying $\eta$ in the  input, one automatically tunes the interplay community \textit{vs} reciprocity: $\eta$ close to 0 gives a network whose edges depend mostly on the community structure imposed by the membership vectors; instead, $\eta$ close to 1 results in a network with lower impact of community structure, i.e.,  reciprocity has also significant impact on the edge formation. Notice that it is not possible to have a contribution purely due to reciprocity, as this phenomenon implicitly requires the existence of another mechanism to produce one of the two possible edges, here the community structure. This can also be seen by observing that Eq.~(\ref{eqn:mij}) can be rewritten as $m_{ij}= \lambda^{0}_{ij}+\f{\eta}{1-\eta^{2}}\, \bup{\eta\, \lambda_{ij}^{0}+\lambda_{ji}^{0}}$; while the first term only depends on communities, the second term depends on both communities and reciprocity and they cannot be separated independently.

Having presented how our model can be used to generate synthetic data, we now proceed in describing how our model relates to observable network properties and how it can be used to predict reciprocated edges.
% ----------------------------------------------------------------------------------------------------------------------------------
\section{Predicting network reciprocity}

In directed networks, reciprocity $ \mathsf{r} $ is usually defined as the fraction of edges that are reciprocated \cite{wasserman1994social}, although other definitions exist to capture this feature \cite{garlaschelli2004, squartini2013reciprocity}.
With our probabilistic model, we can compute the expected value of a related quantity
\be\label{eqn:WRep}
 \mathsf{r}_{w} \coloneqq \f{\sum_{i,j} \rup{A_{ij}\,A_{ji}}}{\sum_{i,j}\rup{A_{ij}}} \quad,
 \ee
  which corresponds to reciprocity in the case of binary adjacency matrices. A natural question is thus how this observable quantity is related to the reciprocity parameter $\eta$. In fact, we show that, provided some assumptions for the second moment $\Exp\rup{A_{ij}^{2}}$ and considering an approximation with Taylor expansion (see \Cref{apxsubsec:expRec}), $\eta$ is a lower bound for it:
\be\label{eqn:expRep}
\Exp\rup{\mathsf{r}_{w}} \approx \eta + \f{\sum_{i,j}\rup{\lambda_{ij}^{0}\, m_{ji} +\eta \, m^{2}_{ji}}}{\sum_{i,j}m_{ij}} \geq \eta \quad .
\ee

The tightness of this bound depends on the latent variables through $\lambda_{ij}^{0}$, (implicitly) $m_{ij}$, and $m_{ji}$. Empirically, we find that in the majority of the experiments the bound is very tight, i.e., $\Exp\rup{\mathsf{r}_{w}} \approx \eta$ and the other terms in Eq.~(\ref{eqn:expRep}) are much smaller than $\eta$ in models with the conditional independence assumption, such as our proposed model with $\eta=0$, where $\Exp\rup{\mathsf{r}_{w}} = \f{\sum_{i,j} m_{ij}\, m_{ji}}{\sum_{i,j}m_{ij}}$. In fact, in these models, the term $\sum_{i,j} m_{ij}\, m_{ji}$ is often very small -- we show empirical evidence of this later. Therefore, even in networks with high reciprocity, models with  conditional independence assumption could poorly reproduce the term. This empirical  result also seems to indicate that the pseudo-likelihood approximation of \Cref{eqn:PSL} is relatively good in our datasets.
The practical indication for practitioners is that networks generated by models with the conditional independence assumption have reciprocity values significantly different from those observed in real data.

% ----------------------------------------------------------------------------------------------------------------------------------
%%%%%%%%%%
\section{Predicting reciprocated edges}
%\label{sec:ed}
\label{sec:edlnkpr}
The dependence structure between pairs of edges should allow us to predict the existence of a reciprocated tie \textit{if} an edge in the opposite direction is observed, such as the citation of a paper if an author has been cited before by someone else. This is a kind of link prediction task, which lets us test the dependence  assumption.  It is also a principled way of comparing the accuracy of various generative models for any real network where no ground truth for the latent variables is known \cite{peel2017ground}.

Conditional edge prediction can be formulated as follows: what is the probability of an edge $i\!\to\!j$ conditioned on observing the opposite  existing edge (or  non-existing edge) $j\!\to\!i$ ? Our model naturally outputs this conditional probability. In contrast, a generative model that assumes conditional independence between edges is not capable of exploiting this extra information. It could only estimate marginal probabilities that do not depend on observing the opposite edge as it uses only the parameters such as community memberships and the affinity matrix. Our model is not capable of fully estimating marginal distributions but nevertheless can estimate its expected value as in Eq.~(\ref{eqn:mij}). This is often the main quantity used in prediction tasks, as it plays the role of a score for estimating the entries $A_{ij}$. Therefore, with our model we can also predict regular edge existence, where we simply aim at predicting an edge without any extra information but the inferred parameters.
%%%%%%%%%%

In our experiments below, we test various generative models for both regular and conditional edge prediction by using 5-fold cross-validation. Specifically, we divide the dataset into five equal-size groups (folds) and give the models access to four groups (training data) for learning the parameters; this contains $80\%$ of the possible pairs of nodes in the network. One then predicts the existence of edges in the held-out group (test set). As performance metrics, we measure the AUC on the test data, i.e., the probability that a randomly selected edge has higher expected value than a randomly selected non-existing edge. We compute both the regular AUC, by using as score the expected value $\Exp_{P(A_{ij}|\Theta)}\rup{A_{ij}}$,  and the \textit{conditional} AUC (AUC$\--$cond), which uses $\Exp_{P(A_{ij}|A_{ji},\Theta)}\rup{A_{ij}}$ as the score, i.e., the expected value over the conditional distribution. The latter can only be computed for our algorithm, as for the others the marginal distribution is the same as the conditional, and thus the two AUC values coincide, see \Cref{appendix:CV} for more details.

\section{Results}
\subsection{Results on real and synthetic data}
We now demonstrate our model by applying it to both real and synthetic data.
In the real-world datasets available to us, we only have a directed network of observed interactions, i.e.,  there is no available ground truth for the actual membership and reciprocity parameters. Consequently, their relative contributions in edge formation cannot be tuned. Thus, we first validate our model and competing algorithms on synthetic data produced with different generative models. We test the  ability of these models to: i) generate sample networks that replicate relevant network quantities such as  reciprocity, similar to the  observed values on the input networks; ii) perform edge prediction tasks.
We then investigate our model's performance on real-world datasets.

In the tests below, we use our model in various ways: the constrained version with constraints on the membership parameters $u$ and $v$ such that $\sum_{k}u_{ik}=\sum_{k}v_{ik}=1,\,\forall i$  (\mtrep), the non constrained version (\mtrepnc), and our model with $\eta = 0$ (\mtrepo), i.e., without considering the reciprocity effect. For comparison, we use two generative models  with latent variables: a community detection-only generative model with a Maximum Likelihood approach \cite{de2017community} (\mt), which was the inspiration for the building block of our model in the case $\eta=0$, and a Bayesian Poisson matrix factorization (\bpmf) commonly used in recommendation systems \cite{gopalan2015scalable}.
For the edge prediction task on real data, we also consider a supervised learning link-prediction routine (\olp) with topological predictors and the implementation of Ghasemian et al. \cite{Ghasemian201914950} (see \Cref{apx:LPf} for details).

\subsection{Performance for synthetic networks}
We study various types of synthetic networks, generated by three different models to cover several network topologies. Two of them cover the extreme scenarios of networks generated, accounting only for community structure or only for reciprocity. For the former we use the standard stochastic block model (SBM) \citep{holland1983stochastic} and for the latter the reciprocity model of Holland and Leinhardt (HL) \citep{holland1981exponential}.
Our model, instead, is designed to tune the relative impact of community structure and reciprocity in determining edges, by varying the parameter $\eta$. Thus,
we use the benchmark generative model described above to interpolate between these two extremes by tuning  $\eta$: for small values we reproduce the results equivalent to the stochastic block model, whereas for higher values we replicate a structure similar to Holland and Leinhardt's model.

The generative process is described in detail in the \Cref{appendix:gen}. As a remark, the exact joint likelihood of \mtrep\text{} is not determined in closed-form, however all the models used here for comparison adopt either its Poisson conditional distribution (our model with $\eta>0$) or its Poisson marginal distribution (all the other models). Thus experiments here are aimed at highlighting differences in the various models' assumptions.  By varying the network sparsity and the impact of communities and reciprocity, we illustrate types of structure that may exist in real-world data, and test each algorithm's robustness against them on various tasks including edge prediction and the ability to reproduce sample networks that replicate relevant network quantities.

\paragraph*{Reproducing the  topological properties}
An important property of a model is the ability to generate network samples that resemble what is observed in real data. We test this ability by considering topological properties like degree distribution, reciprocity, and hierarchical structure. We calculate their values on network samples, which are generated with the various generative models, by applying  the inferred parameters from the given input data. Specifically, we consider networks generated synthetically as explained above, and for each individual network we infer the parameters by each model, and use them to generate five network samples.
We compare topological properties of these samples with those observed on the ground truth networks used to infer the parameters.

In particular, we are interested in measuring reciprocity, as the networks generated by algorithms only based on community structure are not capable of reflecting the observed value of the reciprocity in the ground truth network, a shortcoming of these models which indeed limits their applications. The empirical evidence of this observation was part of the motivation to study this problem.
In the experiments, we use the standard definition of reciprocity $\mathsf{r}$, i.e., the ratio of the number of edges pointing in both directions to the total number of edges in the graph (we use the \verb|python| implementation in \verb|networkx|). As anticipated, in networks generated with the stochastic block model, $\mathsf{r}$ is often close to  $0$. Instead, a more interesting scenario is that of networks generated  with the main purpose of replicating reciprocity, as in the HL model. This is an example of an exponential random graph model where reciprocity and sparsity are the two topological properties controlled in input. It is also one of the few cases where this type of model is analytical, see \Cref{apx:hl}.  In this model, $\mathsf{r}$ is tuned by a parameter  $\alpha$ so that the higher its value, the higher the reciprocity. Notice that, as usual in exponential random graphs models, latent variables such as communities are not considered. This model generates unweighted networks, hence $\mathsf{r} \equiv \mathsf{r}_{w}$.

\begin{figure}[h]
	\includegraphics[width=1\linewidth]{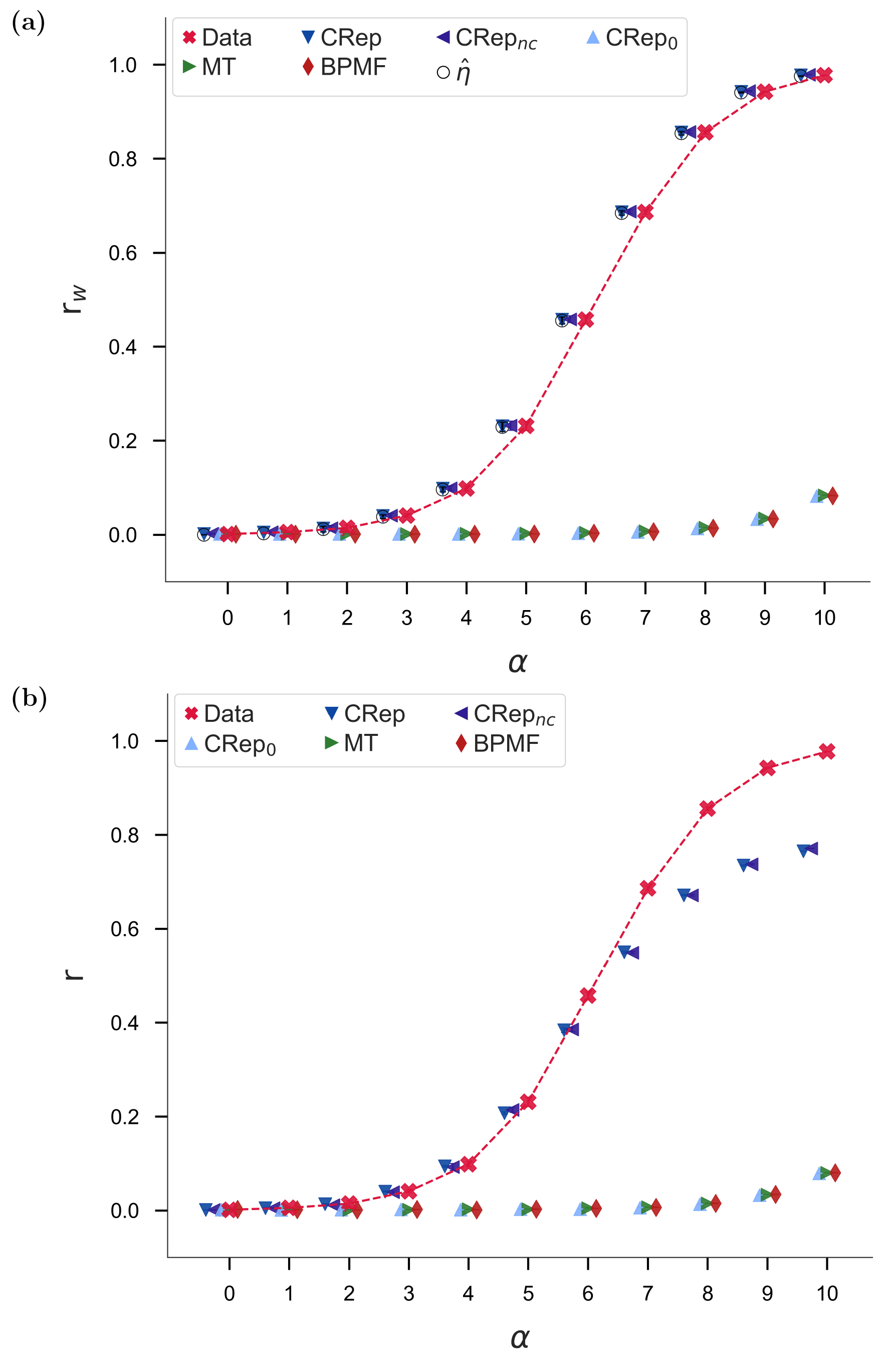}
	\caption{
	\textbf{Reciprocity in HL networks.}  Synthetic networks with $N=1000$ nodes generated with the model proposed by Holland and Leinhardt by varying the reciprocity parameter $\alpha$. Results are empirical averages and standard deviations over 15 samples of three independent synthetic networks (5 samples per input network). The red markers indicate the average on the three input networks. (a) The quantity $\mathsf{r}_{w}$ as defined in Eq. (\ref{eqn:WRep}); the empirical average over the samples and the theoretical expectation as in Eq. (\ref{eqn:expRep}) coincide, hence we omit the markers for the empirical value; $\hat{\eta}$ is the inferred parameter in \mtrep\text{} and \mtrepnc. (b) Standard reciprocity $\mathsf{r}$. Notice that $\mathsf{r}\equiv \mathsf{r}_{w}$ for the input data, but this is not true for the samples, as the generative models considered here generate weighted edges, i.e. the matrix $A$ is in general not binary. Error bars are smaller than marker size. Unless otherwise stated, this will be the case in all of the figures.}
	\label{fig:RecHL}
\end{figure}

Figure~\ref{fig:RecHL} shows that \mtrep \text{} significantly outperforms all the other generative models in reproducing  $\mathsf{r}_{w}$, panel (a), and $\mathsf{r}$, panel (b), as measured on the sampled networks. The gap between the values of $\mathsf{r}$ and $\mathsf{r}_{w}$ on the sampled networks is due to the mismatch between the binary adjacency matrices of the networks generated with the HL model (input data) and the weighted sampled ones generated with the various generative models, which use Poisson distributions. 
%\EPcom{Note that there are extensions of ERGMs that (1) allow for valued edges (see Desmarais and Cranmer PLOS ONE 2012) and (2) are Bayesian (though not with any latent terms, I think) (see Caimo and Friel's work). Presumably you could use the ERGM implementation with weighted edges, instead of the older binary one. I only know of the R package GERGM.}\CDBcom{I would save this only if the reviewers ask.}\MCcom{Interesting. However I agree with CDB since incorporating that means introduce them in the introduction, implementing, and re run all experiment again. Worth it only if a reviewer ask, I guess.} 
Similar results are obtained for the networks generated with our benchmark generative model. Also in this case,  \mtrep\text{} captures reciprocity significantly better than the other models, consistently over a range of values  of $\eta$ as the input parameter. Moreover, in the case of fixed $\eta$,  varying the sparsity  and degree of overlapping communities lead to the  same results.  We leave details in the \Cref{apx:synA}.

%%   --------------------------------------------------------------------------------------------------------------------------------------------------------------------------
%%   --------------------------------------------------------------------------------------------------------------------------------------------------------------------------

At this point, we turn our attention to topological properties other than reciprocity, to investigate how these generative models perform in reproducing various relevant properties that might be of interest for a practitioner. Indeed, other possible mechanisms underlying network interactions are those that involve more than two individuals (which is the case for reciprocity), e.g., hierarchical structure, which requires the whole network for its computation.

As in our experiments we find that all models are able to retrieve the degree distribution with good accuracy,  we mainly focus on replicating ranking of nodes, an application relevant when nodes have a score representing some intrinsic notion of relative strength or prestige. For this, we use SpringRank \cite{debaccoeaar8260}, an algorithm for inferring hierarchies in directed networks that assigns real-valued scores to nodes.  We calculate the Gini index on these scores to provide a global measure for the whole network. Comparing the average over the five samples, we find that \mtrep\text{} and \mtrepo\text{} are able to perfectly retrieve the Gini index of the original network, while the other models tend to overestimate it, see \Cref{apx:synA}. This is consistent over the various synthetic network topologies. Notice that this topological property is  influenced neither by the value of $\eta$, nor the fraction of nodes with mixed-membership used to generate networks; however, it decreases as the average degree, and $\alpha$ increase.

\begin{figure}[h]
	\includegraphics[width=1\linewidth]{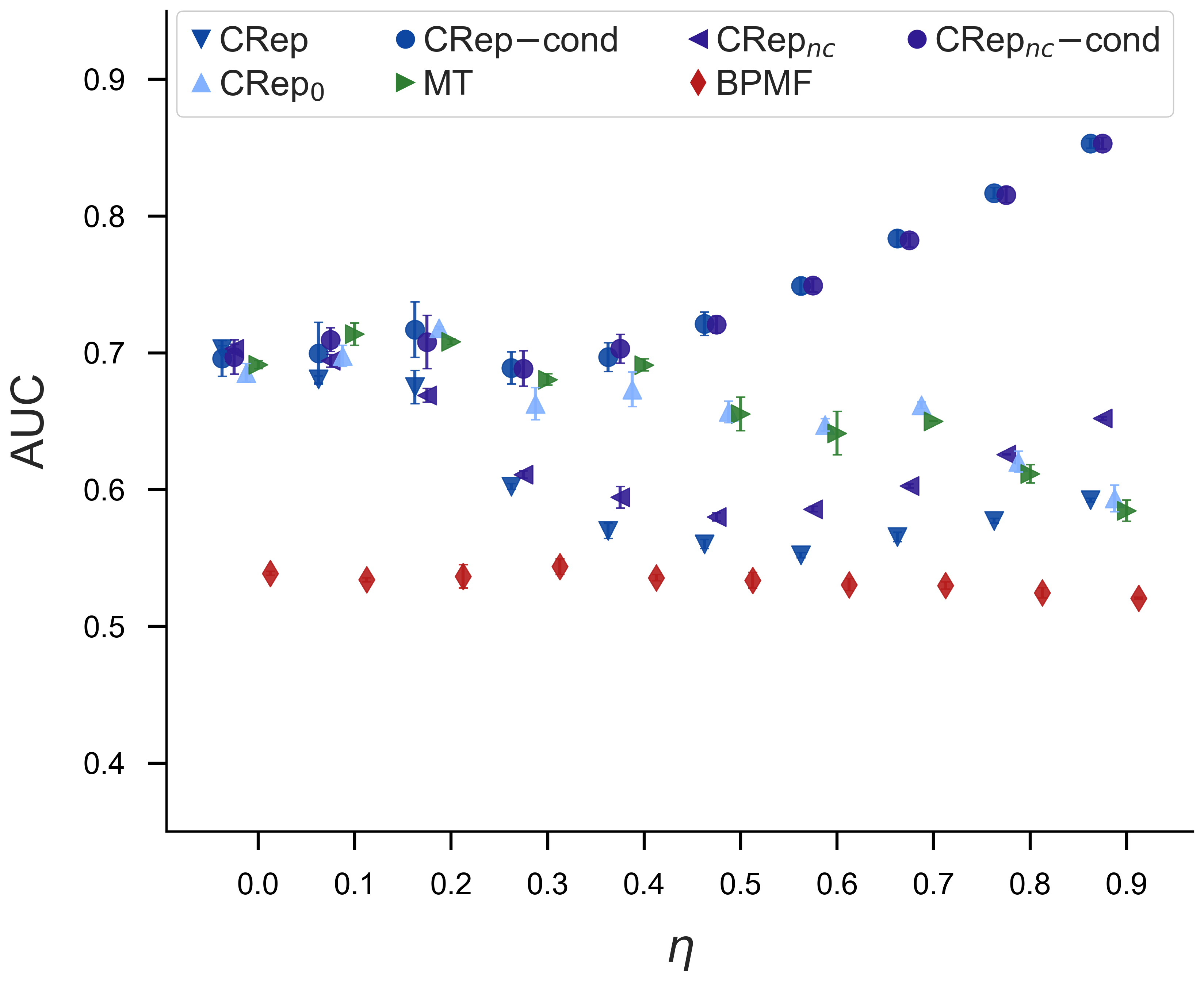}
	\caption{
	\textbf{Edge prediction in benchmark networks.}  Synthetic networks with $N=2100$ nodes and $K=3$ communities of equal-size unmixed group membership generated with the benchmark generative model proposed above by varying the reciprocity parameter $\eta$. The results are averages and standard deviations over three independent synthetic networks and over 5-fold of cross-validation test sets. The accuracy of edge prediction is measured with AUC and the baseline is the random value 0.5.}
	\label{fig:AUCsyn}
\end{figure}

\paragraph*{Edge prediction}
We test the algorithms' ability in edge prediction tasks, in both cases of conditional and regular edge prediction. As we can see from \Cref{fig:AUCsyn},
our model outperforms the others in conditional edge prediction, showing that it is able to efficiently exploit the additional information about the existence of the opposite edge. The performance gap between different approaches increases with $\eta$, as for high values of $\eta$, the reciprocity plays a bigger role in edge formation.  In the opposite scenario of low $\eta$, the impact of reciprocity becomes negligible compared to community structure, and in this case we reproduce the same results as for the other algorithms. This is  expected as our model infers small values of $\eta$ in this case, thus in practice reducing to a conditional independent model as the others.
Performance in terms of regular edge prediction is comparable to the other algorithms for small $\eta$, while it drops for intermediate values and then increases again as $\eta$ grows.

These synthetic tests suggest that working with conditional probabilities results in more robust estimates of the probability that an edge exists if we have access to the edge in the opposite direction. Performance improvement is more significant when community structure is not the predominant mechanism in edge formation. We leave more details in the  \Cref{apx:synB}.

%\HScom{My suggestion is to put the following paragraph in the appendix; it is not necessary to say all the results here.  }
%For sake of completeness, we also validate the model on community detection tasks and observe good performance of \mtrep\text{} in recovering communities when reciprocity has intermediate or low level. Notice, however, that community detection is not the main focus of our model, as we expect the impact of communities in determining the likelihood of an edge to decrease as reciprocity increases, see \Cref{apx:community}. In fact, lower performance in community detection might  signal the presence of other mechanisms for edge formation, like that played by reciprocity. Hence, perfect recovering  of communities should not be expected in these cases.

To summarize results on synthetic networks, \mtrep \text{}  is capable of suitably capturing the reciprocity values observed in a given network, while also retrieving hierarchical structures. Furthermore,  \mtrep\text{} exploits the availability of extra information in performing edge prediction, by increased performance and robustness  across various parameters' ranges.

%%   --------------------------------------------------------------------------------------------------------------------------------------------------------------------------

%   --------------------------------------------------------------------------------------------------------------------------------------------------------------------------

%%%%%%%%%% Real World Data
\subsection{Performance for real networks}
Above, we evaluated the ability of our model, \mtrep \text{}, to generate network samples that have reciprocity values as expected in input and tested its performance in edge prediction.
In this section,  we examine these abilities on real world datasets.
We apply our method to datasets from a diverse set of fields, with sizes ranging up to $N\sim 10^{4}$ nodes and up to $\text{E}\sim 10^{5}$ links (see \Cref{tab:apx_data_desc} and \Cref{apx:data} for details). Together, these examples cover various types of social relationships, communication interactions, transportation systems, and patterns of citations.

\begin{figure}[h]
\includegraphics[width=1\linewidth]{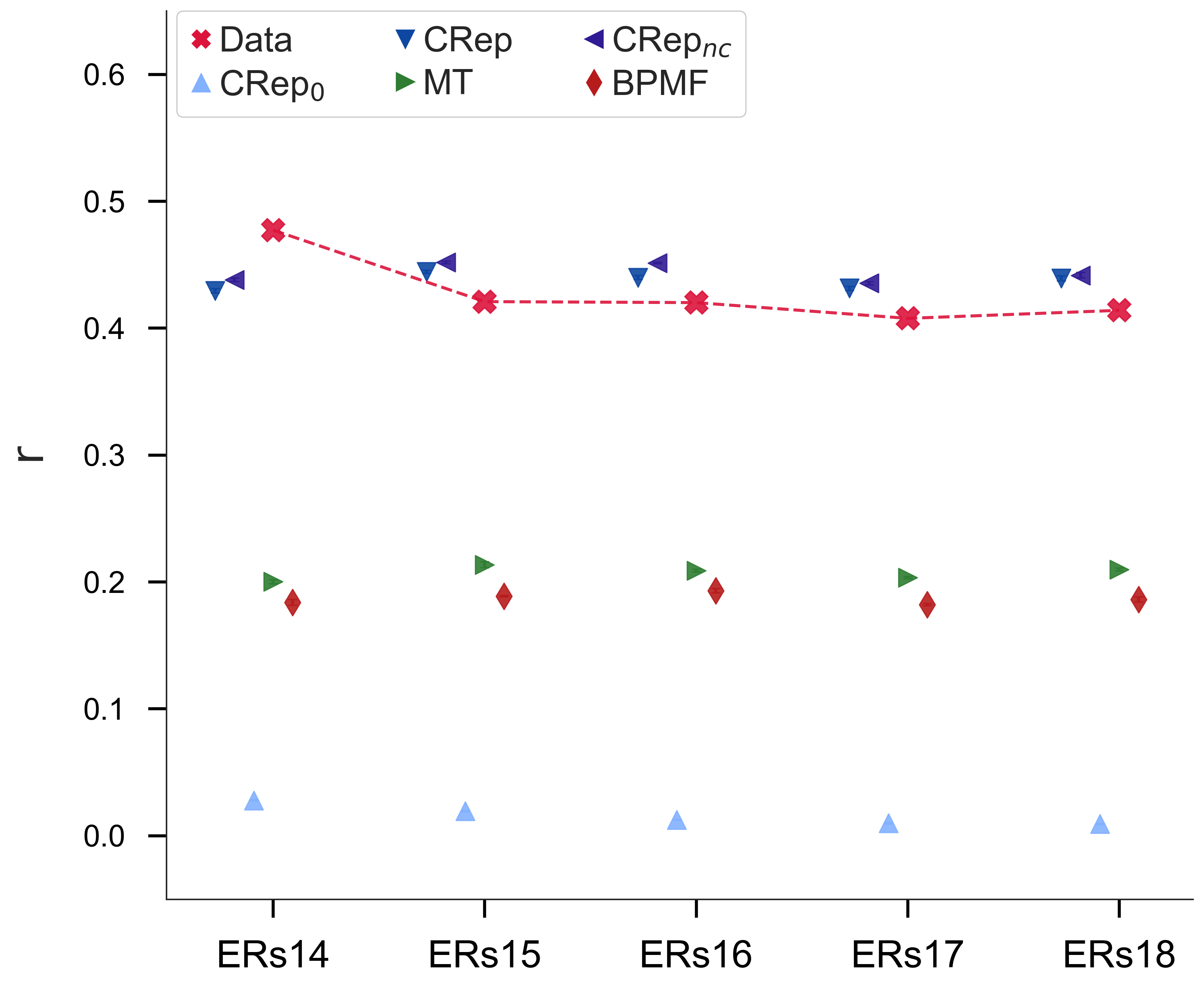}
	\caption{ \textbf{Reciprocity in the Erasmus datasets.} Results are averages and standard deviations of $\mathsf{r}$ over 5 samples generated with the various generative models. The algorithms use the inferred $\eta$ and community parameters of the dataset -- Erasmus in this plot -- to generate synthetic network samples. Red markers indicate the values of $\mathsf{r}$ in the real datasets.}
	\label{fig:feat_Eras}
\end{figure}

\paragraph*{Reproducing the topological properties}
We apply the same procedure as before to infer the parameters $\Theta=(u,v,w,\eta)$ from data (this time, real networks) and then generate synthetic network samples based on them.
Also in this case, \mtrep  \text{} greatly outperforms the other models in  reproducing  $\mathsf{r}$, consistently across datasets. We show as an example in  \Cref{fig:feat_Eras} the results on the Erasmus dataset (Erasmus Mobility Network $2014-2018$) \cite{erasmus}, and we leave the others in the \Cref{apx:realB}.

\begin{figure}[h]
\includegraphics[width=1\linewidth]{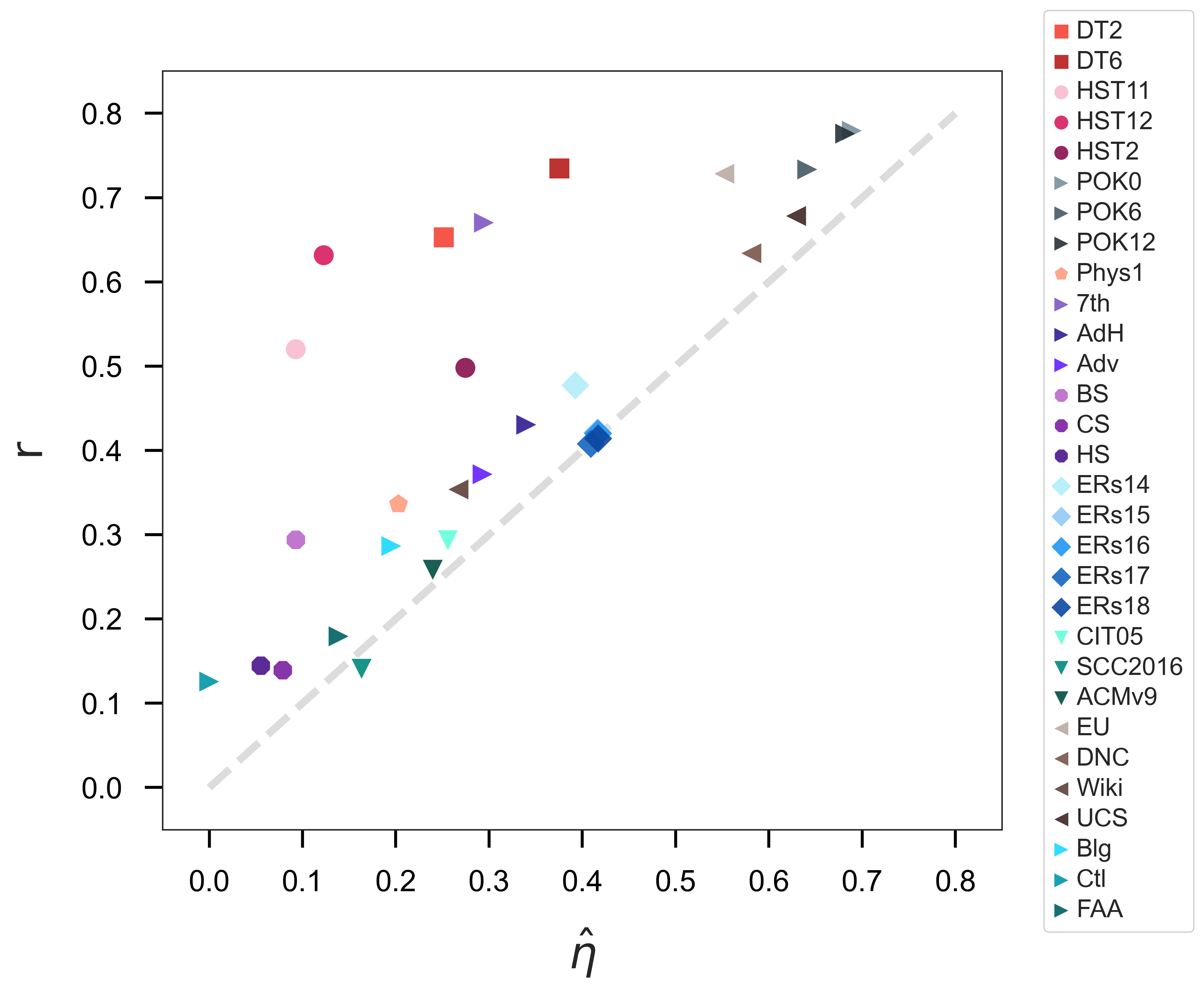}
	\caption{ \textbf{Reciprocity and $\hat{\eta}$.} Scatterplot with observed reciprocity (y-axis) and $\hat{\eta}$  inferred in \mtrep \text{} (x-axis); points are individual real datasets. The dashed grey line indicates the perfect correspondence between $\mathsf{r}$ and $\hat{\eta}$. Marker shape denotes the type of network as defined in \Cref{tab:apx_data_desc}.}
	\label{fig:recEta}
\end{figure}

Previously, we have discussed network-related quantities controlled by $\eta$, such as the expected fraction of edges purely due to communities ($cr_{ratio}$) or the quantity $\mathsf{r}_{w}$. Here we illustrate how the various real networks differ in the inferred values of $\eta$, which we denote as $\hat{\eta}$. In particular, we show in \Cref{fig:recEta} how $\hat{\eta}$ varies according to the reciprocity of these networks, unveiling a non-trivial pattern. While we see a general trend of $\hat{\eta}$ increasing with $\mathsf{r}$, there are  interval ranges of $\mathsf{r}$ for which $\hat{\eta}$ varies widely across networks, and vice-versa. For example, we see that for $\mathsf{r}\in[0.6,0.8]$, $\hat{\eta}$ ranges in $[0.1,0.7]$. This high variability suggests that $\mathsf{r}$ is the result of a complex combination of communities and reciprocity. We notice, for instance, that for high school friendship networks (HST and DT), $\hat{\eta}$ is low (i.e.,  in $[0.1,0.3]$), showing that many reciprocated edges are explained by community structure. Instead, for online dating (POK) and communication networks (EU and DNC), we observe high values of $\hat{\eta}$, signaling a lower impact of communities, as reciprocity plays a bigger role. This reinforces the need to include in network models both mechanisms for explaining edge formation. Notice that these results are possible not only because our model accounts for reciprocity through an explicit parameter $\eta$, but also because it infers reciprocity values  close to the observed ones, while the other methods fail at this, see \Cref{figSI:RecREAL}.
%-------------------------------------------------------------------------------------------------------------------------------------------------------------------

%-------------------------------------------------------------------------------------------------------------------------------------------------------------------
\paragraph*{Edge prediction}
In the absence of ground truth, as in most real world networks, we test the ability in edge prediction by cross-validation, as done for synthetic networks.
\Cref{tab:apx_RD_auc} shows the results in terms of AUC for the generative models \mtrep, \mt \text{},  \bpmf \text{}, as well as for OLP; the latter is a type of supervised learning technique which uses network topological information as features to predict the entries of $A$. \mtrep \text{} and OLP show the best results, with \mtrep \text{} having high performance for social networks. However, if we consider the conditional AUC, then \mtrep\text{} outperforms all the others in the majority of the datasets, as also observed in synthetic data. Finally, by averaging the AUC across the dataset, we find \mtrepnc\text{} is the best model. This confirms the ability of our model to efficiently exploit the additional information from the adjacency matrix to boost performance in terms of edge prediction.

\section{Case study: application of \mtrep \text{} to the Erasmus student exchange network}
\begin{figure*}[htb]
\includegraphics[width=1\linewidth]{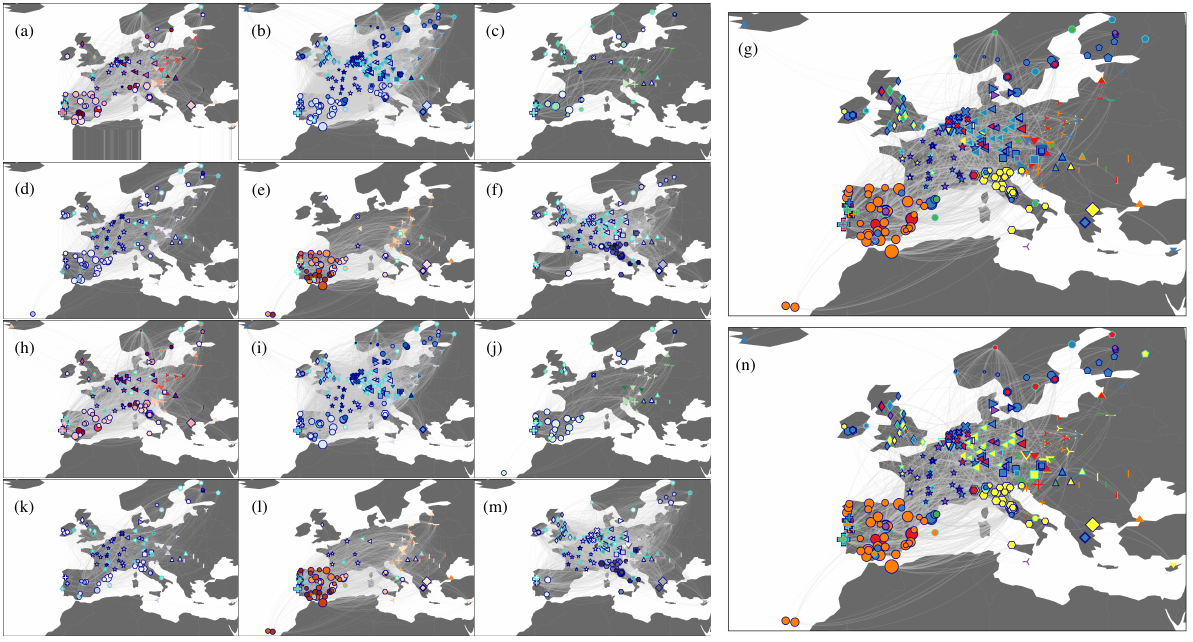}
	\caption{ \textbf{Erasmus 2018 community structure.} For visualization clarity, we show the subnetwork made of the $10\%$ biggest institutions and the 3000 edges with highest weights (inference was performed on the whole network). Panels (a)-(f) show  groups $K=1,2,...,6$ (mixed membership) by \mtrepnc \text{}, and panels (h)-(m) show the same groups by \mt \text{}. Panel (g) illustrates  the groups by  \mtrepnc \text{} in the case of hard membership, while the groups  by \mt \text{} are represented  in panel (n). Node color intensity increases with $u_{ik}$, so that darker nodes have stronger membership $u$ in that group, each color is a group (mixed membership) and nodes with light blue border are nodes that change the most the membership in the two algorithms; for each group $k$, we only show nodes that have $u_{ik}>0.1$.  Node and edge size are proportional to the size of an institution measured by the total number of outgoing and incoming students. Node shapes denote country.}
	\label{fig:QErasmusUV}
\end{figure*}

We illustrate our model on a real dataset to show various analysis that a practitioner can perform. We consider a network representation of the Erasmus student exchange program in 2018 \cite{erasmus}, denoted as ERs18 in \Cref{tab:apx_data_desc}. A node represents a higher education institution  and an edge between nodes $i$ and $j$ denotes how many students were sent from  $i$ to spend a portion of their academic year abroad at institution $j$, as part of their study program towards a degree (Bachelor, Master, or PhD). This program is supported by the European Commission and involves $N=4389$ institutions (mainly European), with a total of $M=90972$ participating students in 2018. \\
We recover community partitions from the network data using both  \mtrepnc \text{} and \mt, they have similar and high performance in edge prediction according to AUC (see \Cref{tab:apx_RD_auc}), and we fix $K=6$ communities from cross-validation. In \Cref{fig:QErasmusUV}, we notice that while both models find several groups that closely  correlate with countries,  \mtrepnc \text{} tends to put German institutions (left triangles) more in the same group (blue) and shifts few
institutions in the red group, which seems made of mainly universities with strengths in engineering and technology (e.g., Universitat Politecnica de Catalunya, Politecnico di Milano and Institut Polytechnique de Grenoble). For instance, Università di Bologna, Federico II di Napoli and Padova have lower $u_{i,red}$ than what is predicted without accounting for reciprocity, instead Slovenská technická univerzita v Bratislave, Kauno Technologijos Universitetas and Universidad de Oviedo increase their membership in this group.

In addition, \mtrepnc \text{} places more institutions with higher membership in the green group, see \Cref{fig:QErasmusUV} (g) (hard membership). While there is no apparent common attribute between these (e.g., country), we find that many nodes with high ``green'' entry of $u_{i}$ tend to reciprocate more edges. Specifically, they have a high fraction of out-neighbors such that $\lambda^{0}_{ij}$ is much smaller than $\lambda^{0}_{ji}$. That is, the edges $A_{ij}$ such that $A_{ji}$ also exists,  have a lower impact in determining the value of $u_{i}$ in the algorithm. In fact  $u_{ik} \propto \sum_{j,q}A_{ij}\rho^{(1)}_{ij}\phi_{ijkq}=\sum_{j,q}\f{A_{ij}\, u_{ik}v_{jq}w_{kq}}{\lambda_{ij}^{0}+\eta\, A_{ji}}$, see Eq.~(\ref{eqn:rho}). Hence, if the denominator is high because of $A_{ji}$, the weight of the edge $A_{ij}$ decreases. Nodes with many such $A_{ij}$ tend to have lower entries $u_{ik}$ and thus lower $\lambda^{0}_{ij}$. This is a qualitative explanation for having different membership, however the situation is more complicated than this, as one needs to account for the effects on the whole network. In fact, also $v_{jq}$ changes between the two algorithms, for a similar reason, thus also contributing to a different $u_{ik}$.

The primary benefit of \mtrep, however, lies not in its ability to recover the communities but in what it reveals about the reciprocity patterns in the network. Home and receiving institutions must sign an inter-institutional agreement to allow for student exchanges between them. While institutions may sign them because of clear affinities between their educational training offerings (e.g., both universities are strong in natural science), they might also do so because of some mechanisms involving reciprocity, as hosting students costs resources. Moreover, reciprocity could be further increased by previous knowledge or collaborations between individual faculties, thus institutional reciprocity may be also driven by faculty reciprocity. In addition to the communities themselves, our model also returns $\eta$, which can reveal features of the data related to such reciprocity effects not seen with standard generative models, such has $cr_{ratio}$ or $\Exp\rup{A_{ij}|A_{ji},\Theta}$. We find a maximum likelihood value of $\eta = 0.4$, signaling a significant reciprocity effect. In fact, according to Eq.~(\ref{eqn:crratio}), on average $40\%$ of the edges are influenced by reciprocity.

%-------------------------------------------------------------------------------------------------------------------------------------------------------------------

%-------------------------------------------------------------------------------------------------------------------------------------------------------------------

\begin{figure*}[htb]
%\captionsetup[subfigure]{justification=centering}
%    \begin{subfigure}[h]{0.5\textwidth}
%    \centering
%       \includegraphics[width=8.5cm]{./figures/OLAVIDE_UV.eps}
%        \caption{Universidad Pablo de Olavide}
%      \end{subfigure}%
%      \hfill
%       \begin{subfigure}[h]{0.5\textwidth}
%          \centering
%       \includegraphics[width=8.5cm]{./figures/MUENCHEN_UV.eps}
%         \caption{Technische Universität München}
%    \end{subfigure}%
%
%        \begin{subfigure}[h]{0.5\textwidth}
%            \centering
%       \includegraphics[width=8.5cm]{./figures/FIRENZE_UV.eps}
%        \caption{Università degli Studi di Firenze}
%            \end{subfigure}%
%            \hfill
%           \begin{subfigure}[h]{0.5\textwidth}
%            \centering
%          \includegraphics[width=8.5cm]{./figures/MADRID_UV.eps}
%          \caption{Universidad Carlos III de Madrid}
%    \end{subfigure}%
\includegraphics[width=0.9\linewidth]{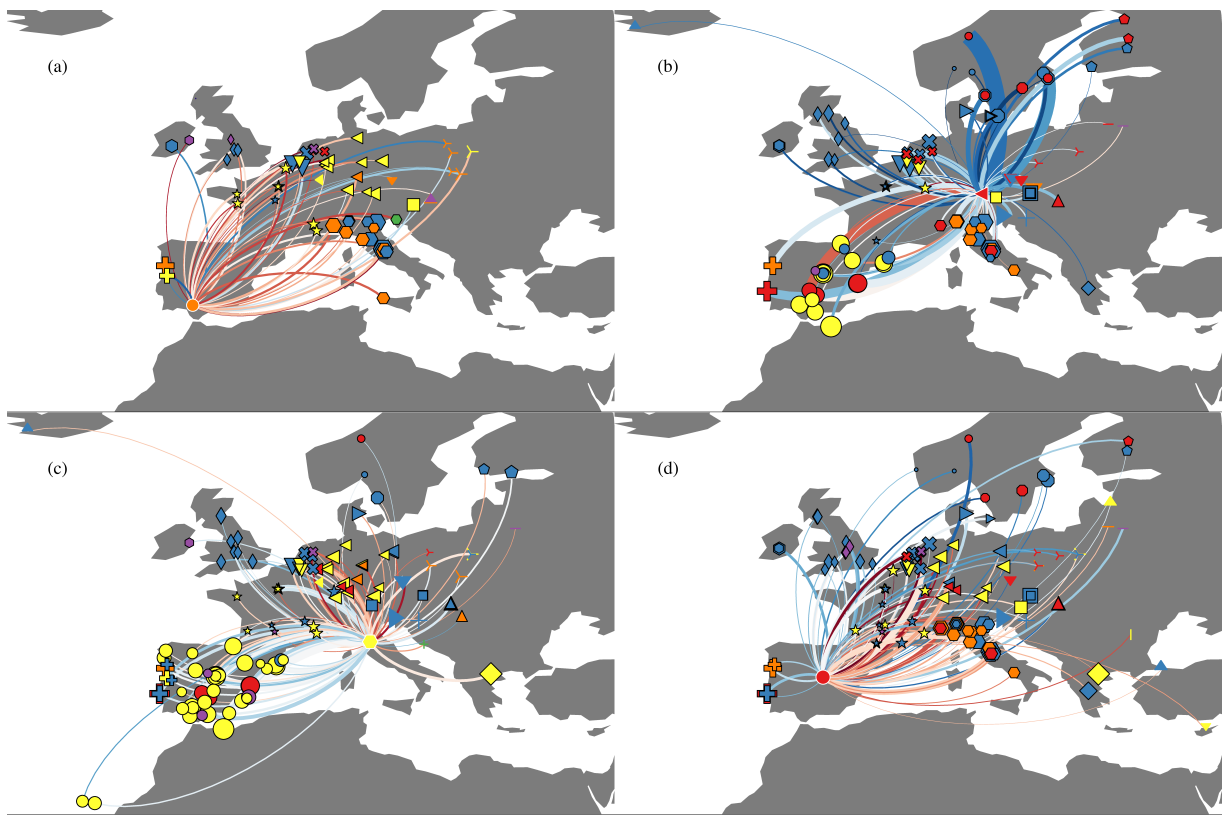}
	\caption{ \textbf{Patterns of reciprocated edges.} Plots show the subnetwork made of the reciprocated edges of (a) Universidad Pablo de Olavide, (b)  Technische Universität München, (c) Università degli Studi di Firenze, and (d) Universidad Carlos III de Madrid. Node size is proportional to university size, the shape denotes country, the colors are the highest entry of $u_{i}$ (for the four reference nodes - white node border) and $v_{j}$ (for all its neighbors). Edge size is proportional to its weight; edge colors vary continuously from red to blue, based on the value of $d_{ij}= cr_{ij}-cr_{ji}$: high intensity red, white and high intensity blue mean  close to -1, 0 and 1 respectively.}
	\label{fig:QErasmusNodes}
\end{figure*}

While $\eta$ gives a global picture of the whole network, our models still allows to distinguish the impact of reciprocity on individual edges. For instance, if an institution $i$ accepts many students from $j$, then $j$ might be more willing to accept students from $i$, even though $i$'s features might not match $j$'s preferences. If we distinguish the $u_{i}$ as the set of preferences of $i$ and $v_{j}$ as the set of attributes of $j$, then our model will naturally convey this through high $\lambda^{0}_{ij}$ and low $\lambda^{0}_{ji}$ for such a case. \mtrep \text{} is able to capture these situations quantitatively, by means of the quantities $cr_{ij}:=\lambda_{ij}^{0}/m_{ij}$ (a $cr_{ratio}$ per edge) with values in $[0,1]$ which measures the relative contribution of communities alone to determine edges between $i$ and $j$. Focusing on a single institution $i$, one can analyze the difference $d_{ij} := cr_{ij}-cr_{ji} \in [-1,1]$ for all $j$ such that both $A_{ij},A_{ji}>0$ and find different reciprocity patterns, as we show in \Cref{fig:QErasmusNodes}. Here we plot three extreme cases where $i$ has most of the $d_{ij}$ being less, equal or greater than 0. The Universidad Pablo de Olavide in Sevilla, panel (a), has mostly $d_{ij}<0$ (plotted in red), meaning that reciprocity has a strong effect in determining its out-going edges to universities that instead send students to Sevilla mostly out of community preference. The opposite case is that of Technische Universität München, panel (b), which has most of the $d_{ij}>0$ (plotted in blue), signaling that it tends to select its out-going edges more out of preference than their counterparts, who tend to reciprocate instead. Università degli Studi di Firenze, panel (c), is an example of an institution with several  $d_{ij}$ close to 0 (plotted in white), meaning that most of its reciprocated edges are due to community affinities. In other words, Firenze selects out-going $j$ based on preference and those who select Firenze do the same, so the impact of reciprocity is low. Apart from these three extremes, many universities display a range of such behaviors; we give an example of Universidad Carlos III de Madrid, panel (d), which has a balanced fraction of reciprocated edges covering these three cases (there are about $1/3$ of blue, red, and white edges in the corresponding figure). Notice that the value of $d_{ij}$ yields an incomplete picture of the situation, since it does not distinguish between cases where the quantities $cr_{ij},cr_{ji}$ have different magnitudes while keeping their difference constant. 

%-------------------------------------------------------------------------------------------------------------------------------------------------------------------
%-------------------------------------------------------------------------------------------------------------------------------------------------------------------

%%%%%%%%%%
\section{Conclusion}
\label{sec:conclusion}
\mtrep \text{} is a mathematically principled generative model for capturing both community and reciprocity patterns in directed networks. It relies on relaxing strict conditional independence assumptions on edges that limit the applicability of standard methods on real problems where reciprocity plays an important role. Its algorithmic implementation is efficient and scalable to large system sizes. The corresponding generative model allows for the creation of synthetic networks with the desired interplay between community and reciprocity in determining the edges, while allowing the tuning of network sparsity.

In addition to providing all the analysis tools typical of standard generative models with communities, our model makes it possible to
 answer questions about reciprocity in networks that were not previously possible; for instance, performing probabilistic conditional edge prediction and estimating the relative contribution of community and reciprocity in determining edges. We show how real networks display a wide range of the reciprocity parameter, signaling the variety of possible patterns for this property.
In the context of the Erasmus student exchange network, our model allowed us to distinguish universities based on their pattern of reciprocated edges.

More generally, our model shows how we can relax strict conditional independence assumptions on edges and showcases possible consequences in doing this. This presents an opportunity for researchers to rethink the fundamental assumptions behind generative models, and present models that may open doors to new theories and questions. We make one step in this direction, as our model connects two popular problems that are mainly treated independently: the inference of communities in networks and generating directed networks where reciprocity plays a relevant role. We used this connection to obtain networks with community structure and values of reciprocity consistent with those observed in real data.

Both the assumption and the model we have presented are only the first step in a broader line of work that investigates how certain topological properties are reflected in networks with latent community structure as dominant mechanism in edge formation. There are a number of directions in which this work could be extended. We have considered here a simple way to account for reciprocity and break conditional independence, by considering a unique  parameter for the whole network.  Our model could be extended to account for node-dependent parameters, where reciprocity varies between individuals. 
In addition, possible extensions may incorporate extra information such as degree, attribute or signals on nodes \cite{contisciani2020,newman2016structure,peel2017ground,hoffmann2020community,stanley2019stochastic}, edges of different types as in multilayer networks \cite{de2017community} and dynamics in time \cite{ghasemian2016detectability,blundell2012modelling,zhang2017random,linderman2014discovering,peixoto2017modelling,mucha2010community}. 
Reciprocity is one of the many effects that could play a role in determining how nodes interact in a network.
One could go further than this by considering incorporating quantities that account for triples of individuals, for instance clustering coefficient, transitivity or global centrality measures \cite{block2015reciprocity}. These properties cannot be captured by standard SBM-like models \cite{seshadhri2020impossibility}.   In this respect, a recent work of Peixoto \cite{peixoto2101disentangling} shares some similarities with ours considering triadic closure instead of reciprocity, making an effort towards extending the stochastic block model framework to incorporate more elaborate topological structure that is not captured otherwise.  This is something that exponential random graphs or stochastic actor oriented models are capable of \cite{robins2007introduction,block2019forms,snijders1996stochastic,snijders2001statistical,snijders2006new}, without including latent community structure but rather fitting network statistics. In probabilistic generative models, this would require further breaking conditional dependencies between edges, potentially increasing the model complexity to encompass more complicated situations. With our work, we made the first step in this direction.

While there is no unique generative model that captures all the possible network properties well, our work illustrates how to target reciprocity. As our original motivation to study this problem came from the realization that standard generative models fail to generate synthetic networks with meaningful values for this property, our work illustrates a way in which latent variable frameworks can be applied more realistically, and provides an example of how network scientists can better align fundamental theories with realistic applications. We provide an open source implementation of the code online at \url{https://github.com/mcontisc/CRep}.

%%%%%%%%%%%%%%%%%%%% Acknowledgements %%%%%%%%%%%%%%%%%%%%
\section*{Acknowledgements}
\vspace{-0.1in}
The authors thank Eleanor Power and Elspeth Ready for useful conversations.
The authors thank the International Max Planck Research School for Intelligent Systems (IMPRS-IS) for supporting Martina Contisciani. 
All the authors were supported by the Cyber Valley Research Fund. 
%\textbf{Author contributions:} All authors derived the model, analyzed results, and wrote the manuscript. 
%\textbf{Competing interests}: The authors declare that they have no competing interests.
%\textbf{Data and materials availability:} All data needed to evaluate the conclusions in the paper are present in the paper and/or the Supplementary Materials. An open source implementation of the code is available online at \url{https://github.com/mcontisc/CRep}.

%%%%%%%%%%%%%%%%%%%% Appendix %%%%%%%%%%%%%%%%%%%%
\appendix
\section*{{Appendix}}
%%%%%%%%%% Synthetic networks %%%%%%%%%%
\section{Synthetic network generation: numerical implementation} \label{appendix:gen}
\vspace{-0.1in}
The synthetic networks used in the analysis are of three types and represent different scenarios: networks with community structure only, with reciprocity only and networks with both communities and reciprocity. In order to obtain networks with only a community structure we use a stochastic block model with different values of average degree $\langle k \rangle$. We generate networks with $K=3$ communities of equal-size unmixed group membership, $N = 2100$ nodes and an assortative structure ($w$ has higher diagonal entries) with main probabilities $p_1 = c \, K / N$ and entries outside the main diagonal equal to $p_2 = 0.1\, p1$, so that the average degree is $\langle k \rangle = c + (K-1)\, c/10$, where $c$ is the average degree within the same community. We generate three independent samples for each value of $c \in [2,20]$, that corresponds to $\langle k \rangle \in [2.4,24]$. On the other hand, we generate networks influenced by reciprocity only through an implementation of the reciprocity model proposed by Holland and Leinhardt (see \Cref{apx:hl}  for details). The input parameter $\alpha$ can be tuned to obtain different values of network reciprocity and we generate three independent samples for each value of $\alpha \in [0, 10]$. We consider $N=1000$ nodes and a probability to generate one of the directed-edges equal to $p=0.002$.

In order to work with synthetic networks having an intrinsic community structure and a given reciprocity value, we use the benchmark generative model proposed in this paper. We generate networks with $N=2100$ nodes and $K=3$ communities by varying three different input parameters: the average degree $\langle k \rangle \in [2, 20]$, the reciprocity coefficient $\eta \in [0, 1)$ and the fraction of nodes with mixed membership $over \in [0, 1]$. While varying one of the parameter, the others are fixed to $\langle k \rangle = 20, \eta = 0.5$ and the degree of overlapping communities $over=0$. In detail, networks are generated in two steps. First, membership vectors $u$ and $v$ are generated following an equal-size unmixed group membership and a Dirichlet distribution with parameter $\alpha=0.1$ for the entries with mixed membership; and the affinity matrix $w$ is generated using an assortative block structures with main probabilities $p_1=K/N$ and secondary probabilities $p_2=0.1\,p_1$. Thus the latent variables $\Theta=(u, v, w, \eta)$ are fixed. Second, edges are drawn according to the generative model described in the main text. Specifically, for each pair of nodes $(i,j)$, i) extract $A_{ij}$ from a Poisson of mean as in Eq.~\eqref{eqn:mij}; ii) extract $A_{ji}$ from a Poisson of mean as in Eq. (\ref{eq:meanPoissonC}). This procedure results in a directed network with the desired reciprocity and sparsity. We generate three independent networks for each value of the three different input parameters.

%%%%%%%%%% Edge Prediction %%%%%%%%%%
\section{Edge prediction and cross-validation}\label{appendix:CV}
%\vspace{-0.1in}
We perform edge prediction using 5-fold cross-validation. In each realization, we divide the dataset, i.e., the entries $A_{ij}$ of the adjacency matrix, into five equal groups selected at random.  We use four of these groups as a training set, to infer the parameters $\Theta$.  We then use the fifth group as a test set, evaluating the score for each $A_{ij}$ in this set, and calculate the AUC value. By varying which group we use as the test set, we get 5 trials per realization. The final AUC is the average over these.
To compute the regular AUC we use as score the expected value $\Exp_{P(A_{ij}|\Theta)}\rup{A_{ij}}=m_{ij}$ as in Eq. (\ref{eqn:mij}); for the \textit{conditional} AUC (AUC$\--$cond), we use as score $\Exp_{P(A_{ij}|A_{ji},\Theta)}\rup{A_{ij}}= \lambda_{ij}^{0}+\eta\, A_{ji}$, i.e.,  the expected value over the conditional distribution. Notice that the latter can only be computed for \mtrep, as for the others $m_{ij}\equiv \lambda_{ij}^{0}$, and thus the two AUC values coincide. The AUC is specified for binary entries, thus the edge weight is not accounted in the evaluation. However, our goal here is to assess edge existence, hence AUC is a suitable metric for this. If a practitioner aims at assessing the quality of the inferred weights as well, then one should specify different metrics for this. \\

%%%%%%%%%% Generative model %%%%%%%%%%
 
\section{Inference: numerical implementation}
All the generative models require inferring $K$, the number of communities. We select this by cross-validation. Specifically, we run several held-out trials  as explained above by varying $K$ and select the value of $K$ that gives the highest (regular) average AUC on the test sets. We then extract the parameters of each method using their best $K$.  For \mt, \bpmf \text{} and \mtrepo, we extract the parameters $u,v,w$; in addition, for \mtrep \text{} and \mtrepnc \text{}, we extract $\eta$. All these algorithms converge to a local optima, as the likelihood landscape is not convex. Hence, we run the algorithm 10 times for different random initializations of the parameters and select the realization that has higher likelihood value.

\section{Detailed derivations}
\label{apx:derivations}
We derive in detail the equations for inferring the parameters. We first apply a variational approach to make the problem tractable, and then use an expectation-maximization algorithm to derive the equations of the updates.
\subsection{Variational approach}\label{apxsubsec:VI}
We aim at maximizing the log-pseudo-likelihood in Eq. (\ref{eqn:LogPSL1}).
The first step is to facilitate the maximization process of the logarithmic term. We consider a probability distribution $\rho_{ij}$ over the two competing terms: this is our estimate of the probability that the edges exist due to the contribution of either the community membership or the reciprocity term.  Applying Jensen's inequality $\log \bar{x} \geq \overline{\log x}$:
\bea
\log \lambda_{ij}&=& \log \left( \rho^{(1)}_{ij} \frac{\lambda^{0}_{ij}}{\rho^{(1)}_{ij }}\,+\, \rho^{(2)}_{ij }\frac{\eta\,A_{ji}}{\rho^{(2)}_{ij}}\right) \nonumber \\
&& \geq  \rho^{(1)}_{ij} \log \frac{\lambda^{0}_{ij}}{\rho^{(1)}_{ij }}\,+ \rho^{(2)}_{ij } \log \frac{\eta\,A_{ji}}{\rho^{(2)}_{ij }} \nonumber \\
 &=&  \rho^{(1)}_{ij} \log {\sum_{k,q} u_{ik} v_{jq} w_{kq}}\,+ \rho^{(2)}_{ij } \log {\eta\,A_{ji}} \nonumber \\
 && - \rho^{(1)}_{ij} \log \rho^{(1)}_{ij }- \rho^{(2)}_{ij } \log \rho^{(2)}_{ij} \quad.
 \label{eq:Jens}
\eea
Moreover, this holds with equality when:
\bea
  \rho^{(1)}_{ij}= \frac{\lambda^{0}_{ij}}{\lambda^{0}_{ij}+\eta\,A_{ji}} \label{eqn:rho1}\quad \text{and} \quad \rho^{(2)}_{ij}= \frac{\eta\,A_{ji}}{\lambda^{0}_{ij}+\eta\,A_{ji}} \label{eqn:rho2} \,.
\eea

\begin{widetext}
Thus maximizing $L^{ps}(\Theta)$ is equivalent to maximizing:
\bea
L^{ps}(\Theta, \rho) &=& \sum_{i,j} \left\{ A_{ij}
  \left( \rho^{(1)}_{ij} \log {\sum_{k,q} u_{ik} v_{jq} w_{kq}}\,+ \rho^{(2)}_{ij } \log {\eta\,A_{ji}} 
   - \rho^{(1)}_{ij} \log{\rho^{(1)}_{ij }}- \rho^{(2)}_{ij } \log{\rho^{(2)}_{ij }} \right)
	-\sum_{k,q} u_{ik}v_{jq}w_{kq} -\,\eta\,A_{ji} \right \} \nonumber \quad.
\label{eq:L_2}
\eea
\end{widetext}

We apply once more the variational approach to make the sum inside the logarithm tractable. Similarly as before, we introduce a probability distribution $\phi_{ijkq} $ such that:
\bea
&& \log{\sum_{k,q} u_{ik} v_{jq} w_{kq}}    \nonumber \\
 && \geq  \sum_{k,q}  \phi _{ijkq} \log {u_{ik} v_{jq}w_{kq}} -\sum_{k,q}  \phi _{ijkq}\log{ \phi_{ijkq}}  \quad.
\eea
The equality holds when:
\begin{equation}
\phi_{ijkq} =\frac{u_{ik}v_{jq}w_{kq} }{\sum_{k',q'}u_{ik'}v_{jq'}w_{k'q'} } = \frac{u_{ik}v_{jq}w_{kq} }{\lambda^{0}_{ij}}\quad.
\label{eqn:phi1}
\end{equation}

\begin{widetext}
Thus maximizing $L^{ps}(\Theta,\rho)$ is equivalent to maximizing:
\bea\label{SIeqn:LPS}
L^{ps}(\Theta, \rho,\phi) &=& \sum_{i,j} \left\{ A_{ij}
 \rho^{(1)}_{ij}\left( \sum_{k,q}  \phi _{ijkq} \log { u_{ik} v_{jq} w_{kq}} - \sum_{k,q}  \phi _{ijkq}\log{ \phi _{ijkq}} \right)+ A_{ij}  \,  \rho^{(2)}_{ij } \log { \eta A_{ji}} \,  \right.  \nonumber \\
 && \left.  \,   -\,A_{ij}  \,  \left(\rho^{(1)}_{ij} \log{\rho^{(1)}_{ij }}+\rho^{(2)}_{ij } \log{\rho^{(2)}_{ij }} \right) -\sum_{k,q}u_{ik}v_{jq}w_{kq}  -\,  \eta A_{ji} \right \} \quad.
\eea
\end{widetext}
with respect to $\Theta$, $\rho$, $\phi$.

\subsection{Expectation-Maximization updates}\label{apxsubsec:EM}
Equations for the updates of each of the parameters can be obtained by taking the derivative of Eq. (\ref{SIeqn:LPS}) with respect to a given parameter and setting it to zero.
For instance, the update equation for $\eta$ is obtained by considering the partial derivative:
\bea
\f{\partial L^{ps}}{\partial \eta} &=& \sum_{i,j} \rup{\f{A_{ij} \rho^{(2)}_{ij}}{\eta } - A_{ji}} \quad.
\eea
Setting this to zero and defining $M= \sum_{i,j} A_{ij}$, we obtain:
\bea
\eta&=&\f{ \sum_{i,j} A_{ij} \rho^{(2)}_{ij}}{\sum_{i,j} A_{ij}} =  \f{ \eta}{M}\,\sum_{i,j}\f{A_{ij}A_{ji} }{\lambda_{ij}}\quad.
\eea
Similarly, for the community affinity matrix we get:
\bea
w_{kq} &=&\frac{\sum_{i,j } A_{ij} \rho^{(1)}_{ij}\phi_{ijkq} }{\sum_{i,j} u_{ik}\, v_{jq}}  \label{eqn:w} \quad .
\eea
%where we used $\rho^{(1)}_{ij}\phi_{ijkq} =\f{u_{ik}v_{jq}w_{kq}}{\lambda_{ij}^{0}+\eta\, A_{ji}}$

Here we show how to enforce constraints like $\sum_{k}u_{ik}=1$, which is an arbitrary choice that can be easily incorporated into our model. To this end, it is convenient to rewrite the log-pseudo-likelihood as follow,
\begin{small}
\bea
 L^{ps}(\Theta, \rho,\phi)&=& F(u_{ik}, v_{jq}, w_{kq})- \sum_{i,j, k,q} u_{ik}\, v_{jq}\, w_{kq}\quad ,\\\nonumber \label{eqn:Ls}
\eea
\end{small}

Then, following the approach in \cite{zhu2013scalable}, to simplify the maximization of the log-pseudo-likelihood, we   substitute $w_{kq}$ from Eq. (\ref{eqn:w}) into Eq. (\ref{eqn:Ls}):
\begin{small}
\bea
 L^{ps}(\Theta, \rho,\phi)  &=& F(u_{ik}, v_{jq}, w_{kq})- \sum_{i,j, k,q} u_{ik}\, v_{jq}\, \frac{\sum_{i,j } A_{ij} \rho^{(1)}_{ij}\phi_{ijkq} }{\sum_{i,j} u_{ik}\, v_{jq}}  \nonumber  \\
  &=&F(u_{ik}, v_{jq}, w_{kq})-\sum_{k,q} \sum_{i,j} u_{ik}\, v_{jq}\, \frac{\sum_{i,j } A_{ij} \rho^{(1)}_{ij}\phi_{ijkq} }{\sum_{i,j} u_{ik}\, v_{jq}}  \nonumber  \\
  &=& F(u_{ik}, v_{jq}, w_{kq})-\sum_{i,j,k,q}   A_{ij} \rho^{(1)}_{ij}\phi_{ijkq} \quad .
\eea
\end{small}
The second term in the above equation does not depend explicitly on $u_{ik}$ and $v_{jq}$.
In order to apply the constraint on the maximization, we add Lagrange multipliers $\gamma_{i}^{u},\gamma_{i}^{v}$:
\begin{widetext}
\bea\label{eqn:LLM}
L^{ps}(\Theta, \rho,\phi) & =&  F(u_{ik}, v_{jq}, w_{kq})-\sum_{k,q}  \sum_{i,j } A_{ij} \rho^{(1)}_{ij}\phi_{ijkq} -\gamma^{u}_i \bup{\sum_{k}u_{ik}-1}-\gamma^{v}_j \bup{\sum_{q}v_{jq}-1}\quad.
\eea
\end{widetext}

The update equation for $u_{ik}$ is obtained by considering the partial derivative,
\bea
\frac{\partial L^{ps}}{\partial u_{ik}} &=& \sum_{j,q } \left( \frac{A_{ij} \rho^{(1)}_{ij}\phi_{ijkq}}{u_{ik}} \right) -\gamma^{u}_i \quad,
\eea
and setting it to zero, which yields:
\bea
u_{ik}&=& \frac{1}{\gamma^{u}_i} \sum_{j,q} A_{ij} \rho^{(1)}_{ij}\phi_{ijkq} \quad.
\eea

By applying the normalization constraint on the $u_{ik}$, i.e., $\sum_{k}u_{ik}=1$, and noticing that $\rho^{(1)}_{ij}\phi_{ijkq} =\f{u_{ik}v_{jq}w_{kq}}{\lambda_{ij}^{0}+\eta\, A_{ji}}$, we can find an expression for $\gamma^{u}_i $:

\bea
\gamma^{u}_i &=& \sum_{j,k,q}\frac{ A_{ij}\, u_{ik}\, v_{jq}\, w_{kq}}{\lambda_{ij}^{0}+\eta\, A_{ji}} =\sum_{j}\frac{ A_{ij}\,\lambda^{0}_{ij}}{\lambda_{ij}^{0}+\eta\, A_{ji}}\quad. \label{eqn:gammau}
\eea
%By using the shorthand notation $M_{ij}=\sum_{k,q}u_{ik}\, v_{jq}\, w_{kq}$, the update equation for $u_{ik}$ is obtained as,
%\bea
%u_{ik}^{(t+1)}&=& \frac{1}{\sum_{j}\frac{ A_{ij}\, M_{ij}}{M_{ij}+\eta\, A_{ji}}} \sum_{j,q} A_{ij} \rho^{(1)}_{ij}\phi_{ijkq} . \label{eqn:uLM}
%\eea

Similarly, we have the following update equation for $v$:
\bea
v_{ik}&=& \frac{1}{\gamma^{v}_i} \sum_{j,q} A_{ji} \rho^{(1)}_{ji}\phi_{jiqk} \quad, \label{eqn:vLM}
%v_{ik}^{(t+1)}&=& \frac{1}{\sum_{j}\frac{ A_{ji}\, M_{ji}}{M_{ji}+\eta\, A_{ij}}} \sum_{j,q} A_{ji} \rho^{(1)}_{ji}\phi_{jiqk} . \label{eqn:vLM}
\eea
where
\bea
\gamma^{v}_i &=& \sum_{j,k,q}\frac{ A_{ji}\, u_{jq}\, v_{ik}\, w_{qk}}{\lambda_{ji}^{0}+\eta\, A_{ij}} =\sum_{j}\frac{ A_{ji}\,\lambda^{0}_{ji}}{\lambda_{ji}^{0}+\eta\, A_{ij}}\quad. \label{eqn:gammav}
\eea

\subsection{Deriving the expected value of the marginal distribution}\label{apxsubsec:meanMarginal} 
\bea
\Exp \rup{A_{ij}} &=& m_{ij} = \sum_{A_{ij},A_{ji}} A_{ij} \, P(A_{ij},A_{ji}|\Theta)  \nonumber \\
 &=& \sum_{A_{ji}} P(A_{ji}|\Theta)\,\sum_{A_{ij}} A_{ij}  \,P(A_{ij}|A_{ji},\Theta) \nonumber \\
&=& \sum_{A_{ji}} P(A_{ji}|\Theta)\, \rup{\lambda^{0}_{ij}+\eta \, A_{ji}} \nonumber \\
&=& \lambda ^{0}_{ij} +\eta \, \sum_{A_{ji}}  A_{ji}\, P(A_{ji}|\Theta) \nonumber \\
&=& \lambda ^{0}_{ij} +\eta \, m_{ji}  \nonumber \\
&=&   \lambda ^{0}_{ij} +\eta \,\bup{\lambda ^{0}_{ji} +\eta \,m_{ij}} \quad.
\eea 
Solving for $m_{ij}$ yields:
\bea
m_{ij } \, \bup{1-\eta^{2}} &=&  \bup{\lambda ^{0}_{ij} + \eta \, \lambda ^{0}_{ji}} \quad,
\eea
which implies:
\be \label{SIeqn:mij}
m_{ij } =\f{ \lambda ^{0}_{ij} +\eta \, \lambda ^{0}_{ji}}{ \bup{1- \eta^{2}} }\quad.
\ee
\subsection{Expected value of $\mathsf{r}_{w}$}\label{apxsubsec:expRec}
With similar calculations as before we obtain:
 \bea
 \Exp \rup{A_{ij}\,A_{ji}} &=& \sum_{A_{ij},A_{ji}}A_{ij}\,A_{ji}\, P(A_{ij},A_{ji}|\Theta)\\
 &=&\lambda_{ij}^{0}\, m_{ji}+\eta \, \Exp\rup{A_{ji}^{2}} \quad.
 \eea
 To fully determine this expression we need to specify the second moment $\Exp\rup{A_{ji}^{2}}$. For binary variables, we could assume $\Exp\rup{A_{ji}^{2}}=\Exp\rup{A_{ji}}=m_{ji}$, as this is the case for Bernoulli distributions. With this assumption, we obtain $ \Exp \rup{A_{ij}\,A_{ji}}=\bup{\lambda_{ij}^{0}+\eta} m_{ji}$. Alternatively, we can assume  $\Exp\rup{A_{ji}^{2}}=m_{ji}+m^{2}_{ji}$ as is the case for the Poisson distribution, and thus obtain $ \Exp \rup{A_{ij}\,A_{ji}}=\bup{\lambda_{ij}^{0}+\eta} m_{ji}+\eta \, m^{2}_{ji}$. Finally we have:
\begin{small} 
% \bea\label{eqnSI:expRep}
%\Exp\rup{ \mathsf{r}_{w} } = \MC{\Exp} \f{\sum_{i,j} \Exp\rup{A_{ij}\,A_{ji}}}{\sum_{i,j}\Exp\rup{A_{ij}}}  &=& \f{\sum_{i,j} \rup{\bup{\lambda_{ij}^{0}+\eta} m_{ji}+\eta \, m^{2}_{ji}}}{\sum_{i,j}m_{ij}}\\
%&=& \eta + \f{\sum_{i,j}\rup{\lambda_{ij}^{0}\, m_{ji} +\eta \, m^{2}_{ji}}}{\sum_{i,j}m_{ij}} \geq \eta \quad. \nonumber
% \eea
 \bea\label{eqnSI:expRep}
\Exp\rup{ \mathsf{r}_{w} } &=& \Exp \rup{\f{\sum_{i,j} \rup{A_{ij}\,A_{ji}}}{\sum_{i,j}\rup{A_{ij}}} } \approx  \f{\sum_{i,j} \nonumber 
\Exp\rup{A_{ij}\,A_{ji}}}{\sum_{i,j}\Exp\rup{A_{ij}}} \nonumber \\
&=& \f{\sum_{i,j} \rup{\bup{\lambda_{ij}^{0}+\eta} m_{ji}+\eta \, m^{2}_{ji}}}{\sum_{i,j}m_{ij}}\nonumber \\
&=& \eta + \f{\sum_{i,j}\rup{\lambda_{ij}^{0}\, m_{ji} +\eta \, m^{2}_{ji}}}{\sum_{i,j}m_{ij}} \geq \eta \quad,
 \eea \end{small}
where in the first row we use the first order Taylor expansion as an approximation. With this assumption, we obtain that the parameter $\eta$ is a lower bound for the expected value of $\mathsf{r}_{w}$.
  An equivalent expression can be derived for models that assume conditional independence, e.g.,  our model with $\eta=0$. In this case we get:
  \begin{small}
  \bea
 \Exp \rup{A_{ij}\,A_{ji}} &=& \sum_{A_{ij}}\, A_{ij}\,P(A_{ij}|\Theta) \sum_{A_{ji}}\,\,A_{ji}\, P(A_{ji}|\Theta) \nonumber\\
 &=&m_{ij}\, m_{ji} \quad,
 \eea
 \end{small}
 which yields:
  \bea
\Exp\rup{\mathsf{r}_{w}} =\f{\sum_{i,j} \Exp\rup{A_{ij}\,A_{ji}}}{\sum_{i,j}\Exp\rup{A_{ij}}} &=& \f{\sum_{i,j} m_{ij}\, m_{ji}}{\sum_{i,j}m_{ij}} \quad.
 \eea

\section{Holland and Leinhardt reciprocity model}
\label{apx:hl}
The model assumes an unweighted and directed network, i.e., asymmetric adjacency matrix with binary values $A_{ij}\in \{0,1\}$, and the following joint probability:
\bea
P(A|\theta,\alpha) &=& \f{e^{-H(A,\theta,\alpha)}}{Z(\theta,\alpha)^{\frac{n(n-1)}{2}}}\\\label{SIeqn:jointHL}
H(A,\theta,\alpha)&=&\theta \sum_{i<j} \bup{A_{ij}+A_{ji}} -\alpha \sum_{i<j}A_{ij}A_{ji} \quad,
\eea
where $Z(\theta, \alpha)=1+2e^{-\theta}+e^{-2\theta +\alpha}$ is the normalization term. The parameter $\alpha$ controls the level of reciprocity, it couples the two entries $A_{ij}$ and $A_{ji}$ thus making the model not factorized; edges between different pairs $(i,j)$ are conditionally independent given the parameters. This is one of the few analytically tractable exponential random graph models. Due to this property, we can extract analytical marginal and conditional distributions for a pair of nodes $(i,j)$:
\bea
P(A_{ij}|\theta,\alpha) &=& \f{e^{-\theta A_{ij} }+e^{-\theta -A_{ij}\bup{\theta-\alpha }}}{Z(\theta,\alpha)}\\
P(A_{ji}|A_{ij},\theta,\alpha) &=&\f{ e^{-A_{ji} \,\bup{\theta-\alpha A_{ij}}}}{1+e^{-\bup{\theta-\alpha A_{ij}}}} \quad.
\eea

These expressions can be used to sample networks with the joint distribution given in Eq. ~(\ref{SIeqn:jointHL}). Tuning the value of the parameter $\alpha$, one generates networks with different values of reciprocity.

\section{Performance in synthetic networks}\label{apx:syn}
\subsection{ Reproducing the  topological properties} \label{apx:synA} 
\begin{figure*}[h]
\includegraphics[width=1\linewidth]{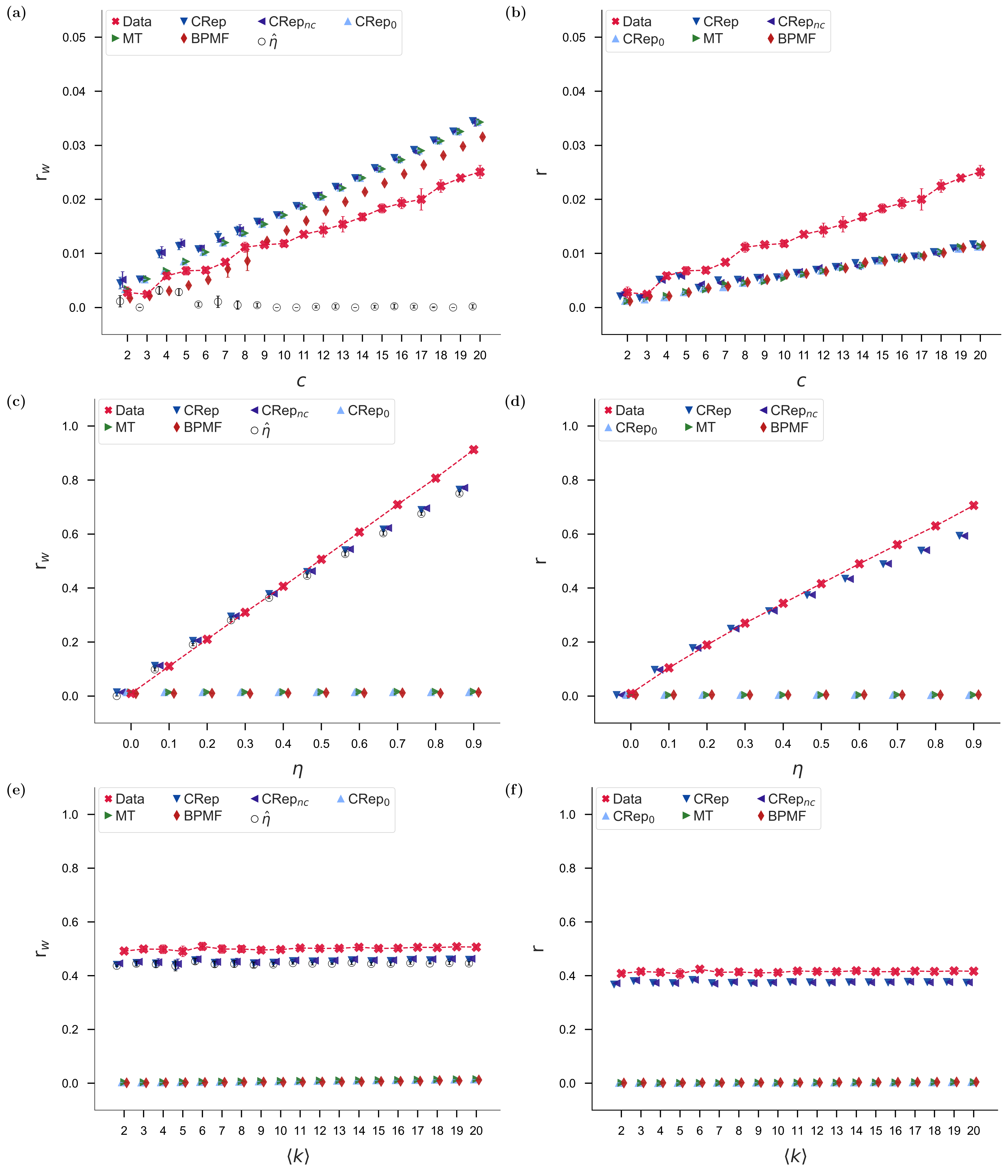}
	\caption{
	\textbf{Reciprocity in synthetic networks}.
	Synthetic networks with $N=2100$ nodes and $K=3$ communities of equal-size unmixed group membership generated with a stochastic block model (a)-(b) by varying the average degree within the same community $c$ and our benchmark generative model, by varying the reciprocity parameter $\eta$ (c)-(d) and the average degree $\langle k \rangle$ (e)-(f).
	Results are empirical averages and standard deviations over 15 samples of three independent synthetic networks (5 sample per input network). The red markers indicate the average on the three input networks. (a),(c),(e) The quantity $\mathsf{r}_{w}$ as defined in Eq. (\ref{eqn:WRep}); $\hat{\eta}$ is the inferred parameter in \mtrep\text{} and \mtrepnc. (b),(d),(f) Standard reciprocity $\mathsf{r}$.  }
\label{figSI:RecSYN}
\end{figure*}

Here we show in more details the ability of the models to reproduce network samples that replicate relevant network quantities. Figure ~\ref{figSI:RecSYN} shows $\mathsf{r}$ and $\mathsf{r}_{w}$ as defined in Eq.~(\ref{eqn:WRep}), computed in the sampled networks of synthetic data generated with a stochastic block model and our benchmark generative model.
As expected, the reciprocity in networks generated with the stochastic block model is always close to zero. Instead, the networks generated with our benchmark generative model present different values of reciprocity, and  \mtrep\text{} captures these values significantly better than the other models, consistently across various magnitudes of input $\eta$. Even in the case of fixed $\eta$, by changing sparsity, we observe the same pattern. By varying the degree of overlapping communities we obtain the same results as changing the average degree (we do not report them here). 

\begin{figure}[p]
\includegraphics[width=1\linewidth]{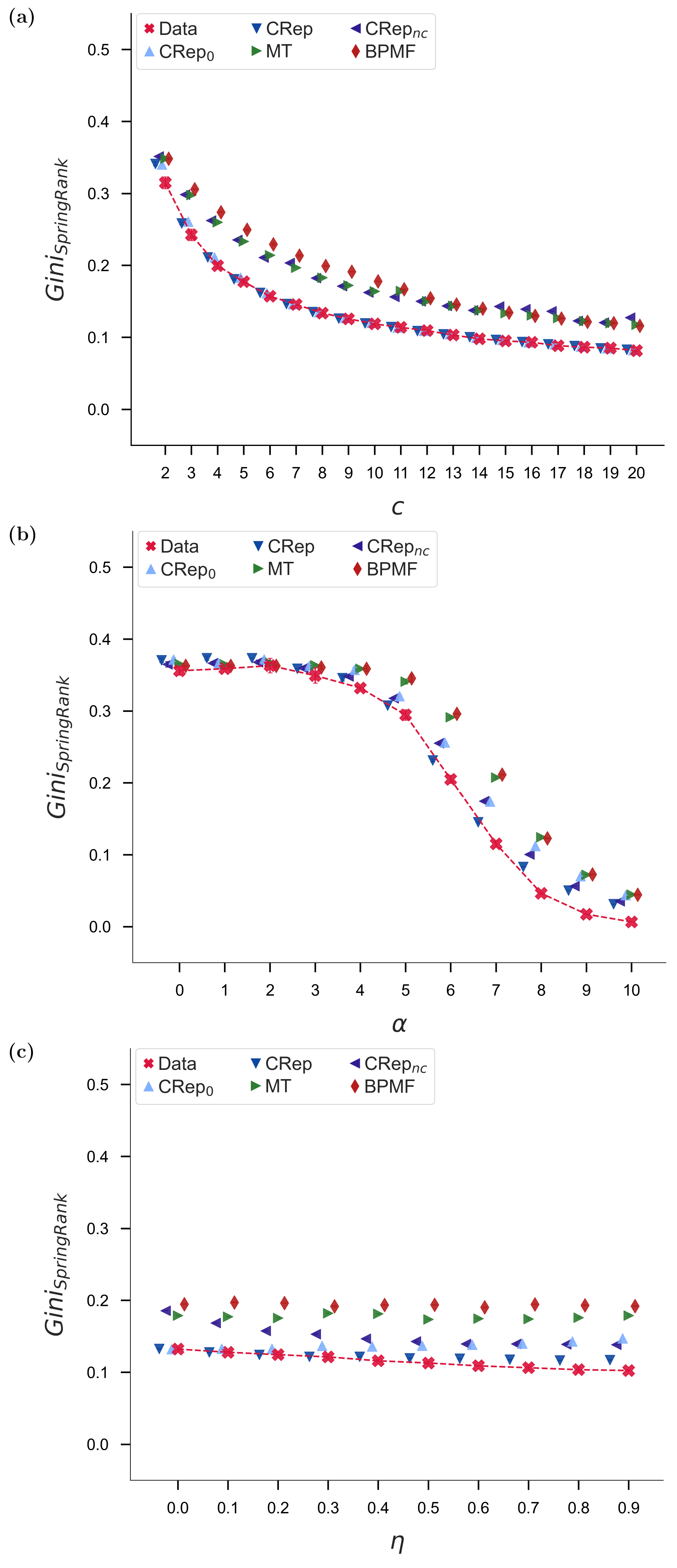}
	\caption{
	\textbf{Hierarchical structure in synthetic networks.}   Synthetic networks generated with (a) the stochastic block model, (b) the HL model, and (c) the benchmark generative model. Results are averages and standard deviations of the Gini index on SpringRank ranking scores over 15 samples of three independent synthetic networks (5 sample per input network). The red markers indicate the average on the three input networks. }
	\label{fig:GiniSYN}
\end{figure}

Figure~\ref{fig:GiniSYN} shows the Gini index computed on nodes scores obtained with the SpringRank algorithm. The Gini index provides a global measure for the whole network, the higher its value, the more hierarchical the network is. We compare the average over the five samples, and we find that \mtrep\text{} and \mtrepo\text{} have reasonable accuracy in retrieving  the Gini index of the original network, while the other models tend to overestimate it. This is consistent over the various synthetic network topologies, i.e., network generated with the stochastic block model, panel (a), the HL model, panel (b), and our benchmark generative model, panel (c). Furthermore, we notice that this topological property decreases as the average degree within the same community, $c$, and $\alpha$ increase, while it is not influenced by the value of $\eta$. We omit the results for the networks generated with our benchmark generative model by varying the sparsity and the fraction of nodes with mixed-membership because we obtain similar results to the stochastic block networks and the benchmark data by varying $\eta$, respectively.

\subsection{Edge prediction in synthetic networks}\label{apx:synB}
\begin{figure}[h]
\includegraphics[width=1\linewidth]{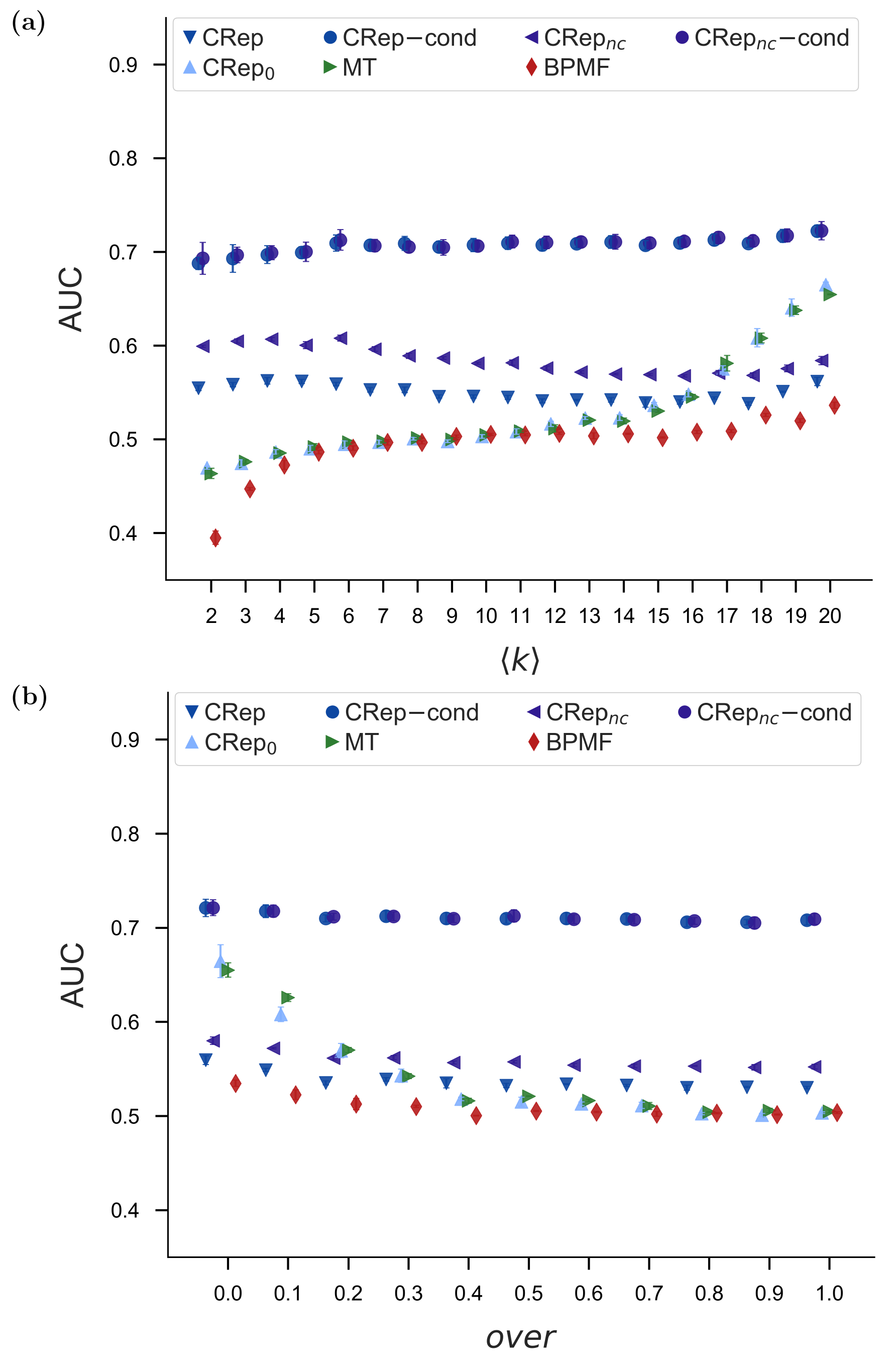}
	\caption{
	\textbf{Edge prediction in synthetic networks.}  Synthetic networks with $N=2100$ nodes and $K=3$ communities of equal-size unmixed group membership generated with the benchmark generative model proposed above by varying (a) the average degree $\langle k \rangle$ and (b) the fraction of nodes with mixed-membership $over$. The results are averages and standard deviations over three independent synthetic networks and over 5-fold cross-validation test sets. The accuracy of edge prediction is measured with AUC and the baseline is the random value $0.5$. }
	\label{fig:AUCkover}
\end{figure}

Here we show the results in terms of edge prediction on synthetic data generated with our benchmark generative model by varying the average degree $\langle k \rangle$ and the fraction of nodes with mixed-membership, which we denote $over$. We use both conditional and regular edge prediction and \Cref{fig:AUCkover} highlights the robustness of \mtrep \text{} and \mtrepnc \text{} in terms of conditional edge predictions, as their performance are significantly higher than that of the other algorithms and do not decrease with increasing overlapping communities and sparsity. Indeed, the results are robust, as we vary the fraction of nodes with overlapping community membership and the average degree, while fixing $\eta=0.5$.  Notice also the stability of \mtrep \text{} and \mtrepnc \text{} in terms of regular edge prediction and how they outperform the other models in critical ranges, e.g., small $\langle k \rangle$ and high $over$.

Moreover, we find more stable results also in terms of regular edge prediction, where \mtrep \text{} and \mtrepnc \text{} have constant values across the different input parameters, outperforming other methods in critical ranges, e.g., small average degree or high overlap between communities. The results of our experiments suggest  that working with conditional probabilities results in more robust estimates of the probability that an edge exists if we have access to the edge in the opposite direction. Performance improvement is more significant when community structure is not the predominant mechanism in edge formation.

\subsection{Community detection in synthetic data}\label{apx:community}
\begin{figure}[h]
\includegraphics[width=1\linewidth]{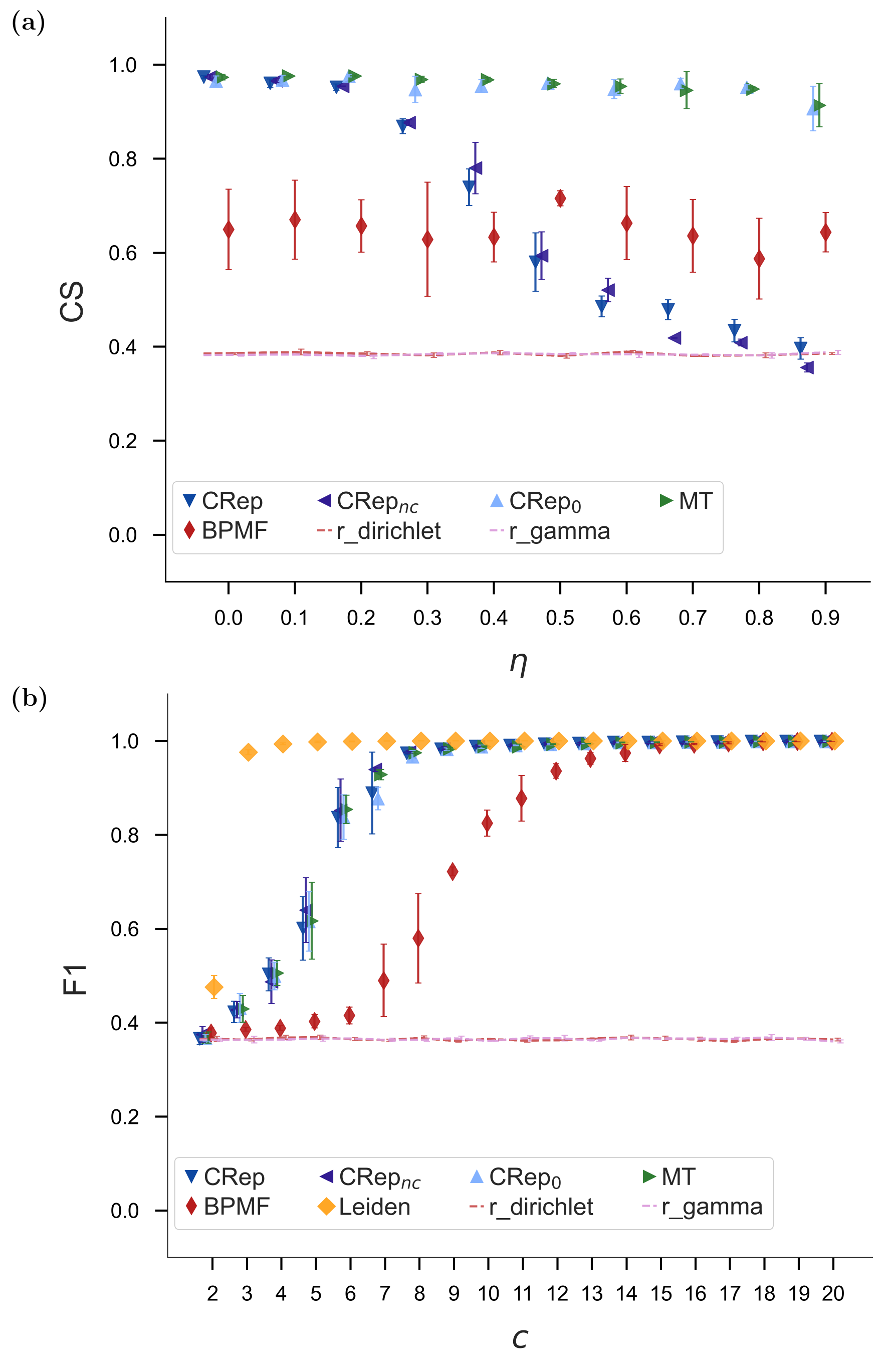}
	\caption{
	\textbf{Community detection in synthetic networks.}  Synthetic networks with $N=2100$ nodes and $K=3$ communities of equal-size unmixed group membership generated with (a) the benchmark generative model proposed above by varying the reciprocity parameter $\eta$, and (b) a stochastic block model. The results are averages and standard deviations over three independent synthetic networks. The accuracy of community detection is measured with (a) cosine similarity and (b) with F1-score as similarity measure, and values close to 1 means higher similarity. The dashed lines represent random baselines, where membership $u_{i}$ are extracted randomly from a Dirichlet of parameter $\alpha=0.1$ or a Gamma distribution of parameters $\alpha=0.1$ and $\beta=1$, to enforce sparsity.}
	\label{fig:CD}
\end{figure}

%%%%%%%%%% Community detection %%%%%%%%%%
For sake of completeness, here we show the performance of the models on  recovering  communities. We consider as performance measure the F1-score (F1) and cosine similarity (CS), the former one is valid for hard membership while the latter captures mixed-membership, we calculate for both the average over the nodes. When measuring the F1-score we consider the entries of maximum value of the membership vectors. Both measures are between $0$ and $1$ and a value of $1$ means perfect reconstruction.  Figure~\ref{fig:CD} shows the accuracy in networks generated with the benchmark generative model by varying the reciprocity parameter $\eta$ and for synthetic data created with a stochastic block model by varying the average degree within the same community $c$. For comparison in these last networks, we consider also the \leiden\text{} algorithm \cite{traag2019louvain}, a non-generative method. Even if  community detection is not the main focus of our model, we notice the ability of \mtrep\text{} in retrieving communities in networks without reciprocity, while its performance decreases as reciprocity increases. This is expected as the community impact in determining the likelihood of an edge decreases as $\eta$ increases. Notice that the benchmark data have been generated with fixed $\langle k \rangle = 20$, thus models without reciprocity are capable of fully  recovering  the community even in the case where reciprocity is there, provided that the average degree is large enough. These synthetic tests suggest, on one side, the robustness of community detection-only methods in  recovering  communities even in the presence of reciprocity; on the other side the good performance of \mtrep \text{} in  recovering  communities when reciprocity has intermediate or low level. This is somehow expected, as this model gives increasingly less weight to the communities as reciprocity increases, thus it is not optimized to recover  the communities when these are not fully determining edge formation.

\section{Performance in real networks}\label{apx:real}

 % ---------------------------------------------------------------------------------------------------------------------------------------------------------
\subsection{Real data: dataset description}
\label{apx:data}
We apply our approach to different types of networks, such as social, ‌ infrastructure, online communication,  and citation networks.
\Cref{tab:apx_data_desc} provides a brief overview of the datasets studied in this work, as well as their abbreviations. All datasets, have been pre-processed as follows: i) self-loops  are removed; ii) only nodes that have at least one out-going and one in-coming edge are kept; iii) we used only the giant connected components.
Some datasets require additional specific pre-processing. Specifically, the citation networks (here: CIT05, SCC2016, ACMv9 ) require extracting a network author-author from a network of paper-citation, so that an edge means that an author cites another author.
Furthermore, we split dynamic networks into separate individual networks where we kept only interactions happening  within a certain time window. This applies to Dutch (DT2, DT6), High school friendships (HST11, HST12, HST2), online dating (POK0, POK6, POK12), and Erasmus (ERs14, ERs15, ERs16, ERs17, ERs18).

% --------------------------------------- Table real data description  --------------------------------------------------------------------------
\begin{table*}[!h]
\Huge
\begin{center}
\caption{{\bf {Datasets description.}}}
\begin{adjustbox}{angle=0}
\resizebox{2\columnwidth}{!}{%
{\renewcommand{\arraystretch}{1.11}
\begin{tabular}{lllllll}
\toprule
 \textbf{Network} & \textbf{Abbreviation}  &\textbf{Category}  & \textbf{N} & \textbf{E}  & \textbf{Ref.}\\
\midrule
 Dutch college & DT2&Human Social Network &$26$&$144$  &  \cite{konect}  \\
 Dutch college & DT6&Human Social Network &$30$&$256$&   \cite{konect}  \\
Highschool Friendships & HST11 &Human Social Network &$31$ &$100$& \cite{konect}    \\
 Highschool Friendships & HST12 &Human Social Network &$30$&$114$ &   \cite{konect}    \\
 Highschool Friendships & HST2 &Human Social Network &$62$&$245$&  \cite{konect}    \\
Online dating        & POK0 &Human Social Network    &$3562$ & $18098$             &    \cite{Makse}    \\
 Online dating        & POK6 &Human Social Network    &$3227$ & $10696$             &  \cite{Makse}    \\
 Online dating        & POK12 &Human Social Network    &$2530$ &$7653$             &  \cite{Makse}    \\
 Physicians             & Phys &Human Social Network    & $95$     & $458$       &   \cite{konect}     \\
 Seventh graders    &7th &Human Social Network       & $29$       &  $376$       &  \cite{konect}     \\
Adolescent health& AdH &Human Social Network      & $2213$   & $11676$     &  \cite{konect}     \\
Advogato              &Adv &Online Social network       & $3858$    &  $42188$   &  \cite{konect}      \\
 Faculty hiring, business department    & BS&Institutions Social Network &$112$  &$3321$& \cite{Clausete1400005}         \\
 Faculty hiring, computer department  & CS&Institutions Social Network &$198$ & $2702$&  \cite{Clausete1400005}         \\
 Faculty hiring, history department      & HS&Institutions Social Network &$140$ &$2242$&   \cite{Clausete1400005}         \\
 Erasmus Mobility Statistics  $2014$   & ERs14 &Institutions Social Network  & $2264$     &$79532$     &  \cite{erasmus}   \\
 Erasmus Mobility Statistics  $2015$   & ERs15 &Institutions Social Network  & $2890$     &$79665$     &   \cite{erasmus}   \\
 Erasmus Mobility Statistics  $2016$   & ERs16 &Institutions Social Network  & $3713$     & $85468$     &  \cite{erasmus}   \\
 Erasmus Mobility Statistics  $2017$   & ERs17 &Institutions Social Network  & $4200$     & $89792$     & \cite{erasmus}   \\
 Erasmus Mobility Statistics  $2018$   & ERs18 &Institutions Social Network  & $4389$     & $90972$     & \cite{erasmus}   \\
 Citation 2005        & CIT05 &Citation Network            & $2130$         & $11153$             &  \cite{snapnets}    \\
 Statistics Citation & SCC2016 &Citation Network       & $2654$   & $21568$       &    \citep{ji2016}    \\
 ACM v9 2012         & ACMv9     &Citation Network      & $8469$         & $56801$            &  \citep{tang2008arnetminer}    \\
Email Eu core network   &EU &Email  Network            & $834$     & $24348$            & \cite{snapnets}    \\
DNC Email             &DNC &Email  Network                 & $548$    & $3575$            &    \cite{konect}    \\
Wiki Talk ht           &Wiki  &Communication  Network & $80$      & $164$            &  \cite{konect}    \\
UC Social               &UCS  &Communication Network & $1302$    & $19044$    &   \cite{konect}    \\
 Blogs &  Blg&Hyperlink Network                              & $830$   & $16107$     & \cite{konect}   \\
Cattle & Ctl&Animal Network                                   &  $24$      & $191$         &  \cite{konect}    \\
FAA Preferred Routes &FAA &Infrastructure  Network   & $1064$   & $2275$       &  \cite{konect}     \\
\bottomrule
\end{tabular}%
}}
\end{adjustbox}
\label{tab:apx_data_desc}
\end{center}
\end{table*}
% ---------------------------------------------------------------------------------------------------------------------------------------------------------

 % ---------------------------------------------------------------------------------------------------------------------------------------------------------
\subsection{Reproducing the topological properties}\label{apx:realB}
\begin{figure*}[!htbp]
	\includegraphics[width=1\linewidth]{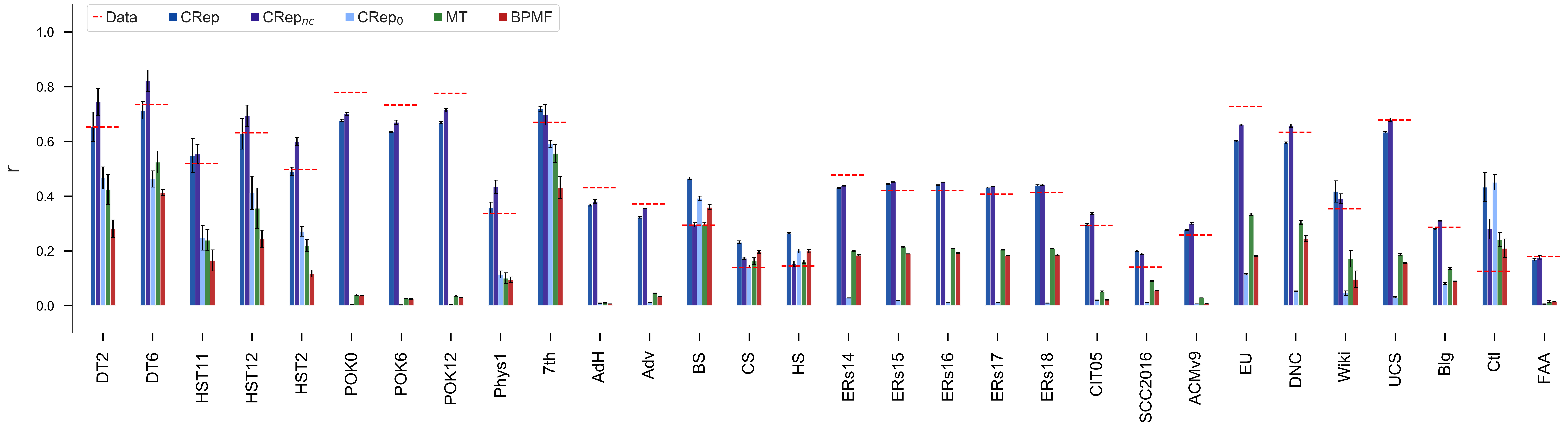}
	\caption{
	\textbf{Reciprocity in real networks}.
	Empirical averages and standard deviations of reciprocity $\mathsf{r}$ over 5 samples of each real network (see \Cref{tab:apx_data_desc} for details). The red dashed lines indicate the $\mathsf{r}$ on the input networks. }
\label{figSI:RecREAL}
\end{figure*}

Here we show the ability of the models to reproduce network samples that replicate relevant network quantities. For each real network we infer the parameters by each model, and use them to generate five synthetic network samples. Figure~\ref{figSI:RecREAL} shows the reciprocity $\mathsf{r}$. For each model, it outputs the averages and the standard deviations over the five samples and the dashed red lines indicate the $\mathsf{r}$ value of the input datasets. We notice the heterogeneity of the analysed networks and how \mtrep\text{} adapts to all different situations, while the other models underestimate the true value most of the times.

\begin{figure*}[!htbp]
	\includegraphics[width=1\linewidth]{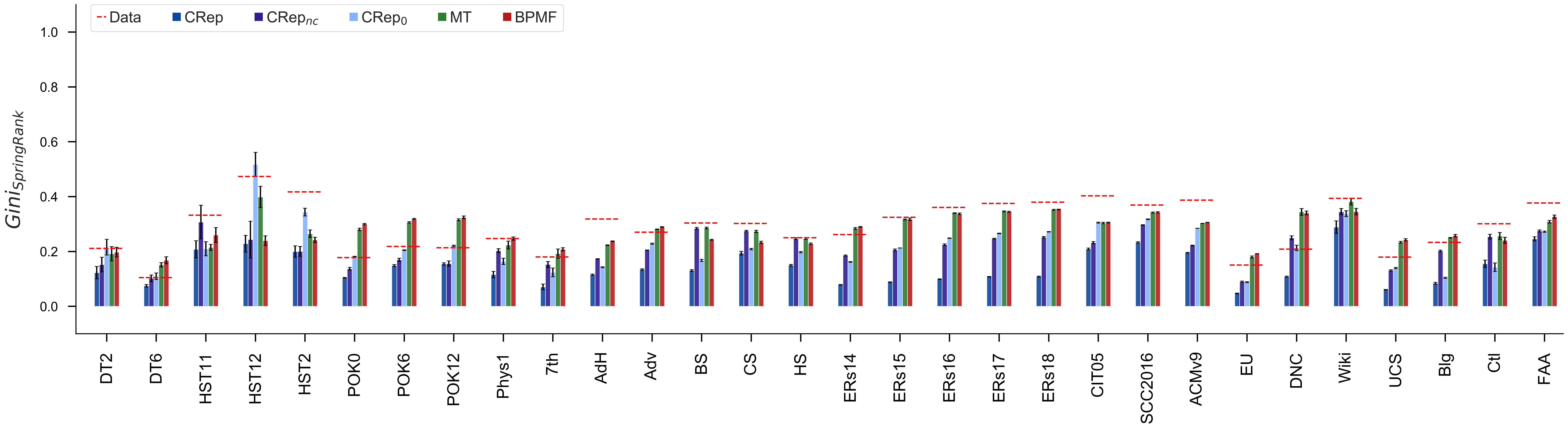}
	\caption{
	\textbf{Hierarchical structure in real networks}.
	Empirical averages and standard deviations of the Gini index on SpringRank ranking scores over 5 samples of each real network (see \Cref{tab:apx_data_desc} for details). The red dashed lines indicate the values on the input networks. }
\label{figSI:GiniCcREAL}
\end{figure*}

Figure~\ref{figSI:GiniCcREAL} shows the Gini index computed on nodes scores obtained with the SpringRank algorithm. The results vary widely depending on the datasets, and we cannot draw general conclusions. In this scenario, we have also studied the reproducibility of the clustering coefficient, i.e., the tendency of nodes to form edges within the same neighborhood, however, we obtain poor results in line with the SBM approach, as predicted in \cite{seshadhri2020impossibility}. Moreover, these are topological properties that involve more complex interactions than pairwise, as in the case of reciprocity (clustering involves triangles and SpringRank score is a global measure). This suggests that, in order to have better performance, one would need to develop more complex models, for instance extending the ideas behind \mtrep \text{} to capture triadic interactions, possibly guided by domain-knowledge about how triadic interactions and reciprocity are related \cite{block2015reciprocity}. We leave this for future work, noting that while exponential random graph models can do this, they do not include latent community structure (analogously as for reciprocity).

 % ---------------------------------------------------------------------------------------------------------------------------------------------------------
\subsection{Link prediction features}
\label{apx:LPf}

Here we present the supervised learning-link prediction routine (OLP) used for comparison in the edge prediction task on real data. In the link prediction task, scores are assigned to all possible pairs of nodes in the graph based on a set of criteria. Then, the pairs of nodes are sorted according to their scores in an ascending order and the most-likely links are the pairs with scores above a threshold value.

Two categories of features are used to determine the criteria of link classification: (i) global features, defined based on the features of the entire network, such as the number of nodes, number of edges, average degree of nodes, and the average clustering coefficient, and (ii) local features, which include the descriptive features of a single node or a pair of nodes.

In this work, we apply the extended definition of features for a directed network of Ghasemian et al. \citep{Ghasemian201914950}.
We also examine  the effect of belonging to the same community  on the local pairwise features,  i.e., pairwise  attributes contribute  in the link prediction only if the two nodes belong to the same community.
However, we did not find significant changes and at the price of higher computational cost, hence, we exclude this factor from the study and omit the results.
Considering $\Gamma(x)_{out/in}$ as the set of  out/in-neighbors of node $x$, and $d(x,y)$ as the distance between nodes $x$ and $y$, some of the well-known features deployed for link prediction are presented in \Cref{tab:lp}.

% --------------------------------------- Table link prediction features  --------------------------------------------------------------------------
\begin{table*}[!htbp]
\Huge
\begin{center}
\caption{\bf {Extended features used in the link prediction process for a directed network. }}
\label{tab:lp}
\begin{adjustbox}{angle=0}
\resizebox{2\columnwidth}{!}{%
{\renewcommand{\arraystretch}{1.11}
\begin{tabular}{llll}
\toprule
& \bf{Feature}	& \bf{Description}	 \\
\midrule
&\textbf{Common neighbors	 out/in}	& defined for a pair of nodes: $x,y$: $\lvert \Gamma(x)_{out/in}  \cap \Gamma(y)_{out/in}\rvert$		 \\
&\textbf{Jaccard index} 	&  defined for a pair of nodes: $x,y$: $\frac{\lvert \Gamma(x)_{out/in}  \cap \Gamma(y)_{out/in}\rvert}{\lvert \Gamma(x)_{out/in}  \cup \Gamma(y)_{out/in} \rvert}$	     \\
&\textbf{Adamic–Adar index}	& defined for a pair of nodes: $x,y$: $\sum_{z \in \{\Gamma(x)_{out/in}   \cap \Gamma(y)_{out/in}\} } \frac{1} {\log\lvert \Gamma(z)\rvert}$		 \\
&\textbf{Resource Allocation index} 	 & defined for a pair of nodes: $x,y$: $\sum_{z \in \{\Gamma(x)_{out/in} \cap \Gamma(y)_{out/in}\}}\frac{1} {\lvert \Gamma(z)\rvert}$		 \\
&\textbf{Betweenness centrality }	& a measure of node centrality based on the shortest paths \\
&\textbf{Closeness centrality} 	& defined for a pair of nodes: $x,y$: $\frac{1} { \sum_{y  } d(y,x)}$		 \\
&\textbf{Shortest Paths} 	& shortest path between nodes: $x,y$	 \\
&\textbf{Katz centralities} 	&  a measure of centrality in a network \\
&\textbf{PageRank centralities} 		&  a measure of the importance of a node  as an  adjustment of Katz centrality	 \\
&\textbf{Eigenvector centralities} 	& an adjustment of Katz centrality of a node in regards to the importance of its neighbors   	 \\
&\textbf{Clustering coefficient for node $x$} 	& $\frac{\text{number of triangles connected to node}\, x}{\text{number of triples centered around node}\, x}$ \\
&\textbf{Preferential attachment} 	&  the tendency of nodes to connect to the nodes with higher degree	 \\
&\textbf{Common community}	&  1 if the pair of nodes belong to the same community, otherwise zero 	 \\
\bottomrule
\end{tabular}
}}
\end{adjustbox}
\end{center}
\end{table*}

 % ---------------------------------------------------------------------------------------------------------------------------------------------------------

\begin{table*}[!htbp]
\Huge
\caption{{\bf Edge prediction in real networks.} Regular AUC and conditional AUC (AUC$\--$cond) for all real networks (see \Cref{tab:apx_data_desc} for details). Results are averages and standard deviations over 5-fold cross-validation test sets. In grey box we show the best performance over all methods, while in boldface the best results in terms of regular AUC. The last row  reports the average and standard deviation of each method over datasets. }
\begin{adjustbox}{angle=0}
\resizebox{2\columnwidth}{!}{%
\begin{tabular}{lllllll | ll}
%\toprule

\toprule
\multicolumn{1}{l}{} &
\multicolumn{6}{c}{\bf{AUC}}    &
\multicolumn{2}{c}{\bf{AUC$\--$cond}}    \\
\cmidrule(lr){2-7}
\cmidrule(lr){8-9}

 \bf{Dataset} &                     \bf{ CRep} &           \bf{CRep$_{nc}$} &               \bf{ CRep0 }&                   \bf{MT }&                                              \bf{BMPF} &                                   \bf{OLP} &    \bf{CRep}&        \bf{CRep$_{nc}$} \\
\midrule
     DT2 &           0.71 $\pm$ 0.01 &      \textbf{0.73 $\pm$ 0.01} &    0.653 $\pm$ 0.009 &      0.71 $\pm$ 0.03 &                                   0.72 $\pm$ 0.01 &                                 0.712 &                        0.77 $\pm$ 0.02 &      \colorbox{lightgray}{0.79 $\pm$ 0.03} \\
     DT6 &           0.72 $\pm$ 0.03 &               0.76 $\pm$ 0.01 &      0.72 $\pm$ 0.01 &    0.762 $\pm$ 0.006 &                        \textbf{0.774 $\pm$ 0.008} &                                 0.737 &                        0.83 $\pm$ 0.03 &      \colorbox{lightgray}{0.85 $\pm$ 0.02} \\
   HST11 &  \textbf{0.74 $\pm$ 0.01} &               0.73 $\pm$ 0.01 &      0.63 $\pm$ 0.03 &      0.62 $\pm$ 0.03 &                                   0.63 $\pm$ 0.04 &                                 0.714 &  \colorbox{lightgray}{0.78 $\pm$ 0.02} &                            0.76 $\pm$ 0.02 \\
   HST12 &  \textbf{0.82 $\pm$ 0.02} &             0.801 $\pm$ 0.008 &    0.743 $\pm$ 0.004 &      0.74 $\pm$ 0.01 &                                   0.76 $\pm$ 0.02 &                                 0.778 &                        0.85 $\pm$ 0.01 &      \colorbox{lightgray}{0.86 $\pm$ 0.02} \\
    HST2 &         0.771 $\pm$ 0.009 &               0.76 $\pm$ 0.01 &      0.73 $\pm$ 0.01 &      0.73 $\pm$ 0.01 &                                   0.71 $\pm$ 0.01 &  \colorbox{lightgray}{\textbf{0.828}} &                      0.808 $\pm$ 0.009 &                            0.79 $\pm$ 0.02 \\
    POK0 &       0.7747 $\pm$ 0.0001 &    \textbf{0.845 $\pm$ 0.002} &    0.665 $\pm$ 0.002 &    0.7400 $\pm$ 0.0009 &                               0.7652 $\pm$ 0.0002 &                                 0.804 &                      0.908 $\pm$ 0.002 &    \colorbox{lightgray}{0.934 $\pm$ 0.002} \\
    POK6 &         0.758 $\pm$ 0.001 &    \textbf{0.818 $\pm$ 0.002} &    0.587 $\pm$ 0.003 &    0.626 $\pm$ 0.002 &                               0.6939 $\pm$ 0.0007 &                                  0.750&                      0.884 $\pm$ 0.005 &    \colorbox{lightgray}{0.909 $\pm$ 0.002} \\
   POK12 &         0.765 $\pm$ 0.002 &    \textbf{0.833 $\pm$ 0.002} &    0.582 $\pm$ 0.002 &    0.606 $\pm$ 0.002 &                               0.6723 $\pm$ 0.0006 &                                 0.739 &                      0.905 $\pm$ 0.003 &    \colorbox{lightgray}{0.924 $\pm$ 0.002} \\
   Phys1 &           0.600 $\pm$ 0.008 &    \textbf{0.627 $\pm$ 0.006} &    0.556 $\pm$ 0.009 &      0.57 $\pm$ 0.01 &                                    0.60 $\pm$ 0.02 &                                 0.577 &                      0.676 $\pm$ 0.005 &      \colorbox{lightgray}{0.71 $\pm$ 0.01} \\
     7th &           0.69 $\pm$ 0.02 &               0.79 $\pm$ 0.01 &      0.72 $\pm$ 0.02 &      0.800 $\pm$ 0.009 &                        \textbf{0.809 $\pm$ 0.005} &                                 0.494 &                        0.77 $\pm$ 0.01 &      \colorbox{lightgray}{0.84 $\pm$ 0.01} \\
     AdH &         0.678 $\pm$ 0.003 &             0.696 $\pm$ 0.002 &    0.656 $\pm$ 0.002 &    0.666 $\pm$ 0.003 &                                 0.627 $\pm$ 0.004 &  \colorbox{lightgray}{\textbf{0.867}} &                       0.760 $\pm$ 0.003 &                          0.787 $\pm$ 0.001 \\
     Adv &         0.771 $\pm$ 0.002 &           0.8919 $\pm$ 0.0001 &     0.760 $\pm$ 0.003 &    0.887 $\pm$ 0.001 &                               0.8907 $\pm$ 0.0005 &   \colorbox{lightgray}{\textbf{0.940}} &                       0.830 $\pm$ 0.002 &                        0.9333 $\pm$ 0.0005 \\
      BS &         0.662 $\pm$ 0.004 &  \textbf{0.8749 $\pm$ 0.0006} &    0.649 $\pm$ 0.004 &  0.8749 $\pm$ 0.0005 &                               0.8746 $\pm$ 0.0009 &                                 0.711 &                        0.66 $\pm$ 0.01 &   \colorbox{lightgray}{0.8750 $\pm$ 0.0006} \\
      CS &         0.715 $\pm$ 0.008 &             0.829 $\pm$ 0.001 &    0.696 $\pm$ 0.005 &     0.830 $\pm$ 0.002 &                                 0.838 $\pm$ 0.001 &  \colorbox{lightgray}{\textbf{0.844}} &                      0.709 $\pm$ 0.008 &                          0.833 $\pm$ 0.001 \\
      HS &         0.661 $\pm$ 0.005 &             0.866 $\pm$ 0.003 &    0.646 $\pm$ 0.003 &    0.866 $\pm$ 0.003 &  \colorbox{lightgray}{\textbf{0.872 $\pm$ 0.001}} &                                 0.865 &                      0.654 $\pm$ 0.005 &                          0.867 $\pm$ 0.003 \\
   ERs14 &         0.754 $\pm$ 0.001 &  \textbf{0.9157 $\pm$ 0.0005} &    0.696 $\pm$ 0.009 &  0.9115 $\pm$ 0.0004 &                               0.9123 $\pm$ 0.0003 &                                 0.893 &                       0.810 $\pm$ 0.001 &  \colorbox{lightgray}{0.9278 $\pm$ 0.0002} \\
   ERs15 &           0.79 $\pm$ 0.01 &  \textbf{0.9361 $\pm$ 0.0002} &      0.72 $\pm$ 0.02 &   0.9330 $\pm$ 0.0002 &                               0.9312 $\pm$ 0.0002 &                                 0.929 &                        0.82 $\pm$ 0.01 &  \colorbox{lightgray}{0.9454 $\pm$ 0.0002} \\
   ERs16 &       0.8057 $\pm$ 0.0006 &  \textbf{0.9454 $\pm$ 0.0002} &  0.7064 $\pm$ 0.0004 &  0.9402 $\pm$ 0.0003 &                               0.9419 $\pm$ 0.0001 &                                 0.944 &                    0.8346 $\pm$ 0.0006 &  \colorbox{lightgray}{0.9552 $\pm$ 0.0002} \\
   ERs17 &         0.822 $\pm$ 0.005 &           0.9484 $\pm$ 0.0001 &    0.734 $\pm$ 0.002 &  0.9433 $\pm$ 0.0002 &                               0.9468 $\pm$ 0.0002 &                         \textbf{0.950} &                      0.838 $\pm$ 0.005 &  \colorbox{lightgray}{0.9568 $\pm$ 0.0002} \\
   ERs18 &       0.8334 $\pm$ 0.0006 &           0.9501 $\pm$ 0.0001 &    0.732 $\pm$ 0.002 &  0.9444 $\pm$ 0.0002 &                                0.9490 $\pm$ 0.0002 &                        \textbf{0.952} &                    0.8476 $\pm$ 0.0006 &  \colorbox{lightgray}{0.9579 $\pm$ 0.0001} \\
   CIT05 &          0.910 $\pm$ 0.002 &           0.9189 $\pm$ 0.0008 &    0.901 $\pm$ 0.001 &    0.918 $\pm$ 0.001 &                                 0.908 $\pm$ 0.001 &  \colorbox{lightgray}{\textbf{0.954}} &                      0.928 $\pm$ 0.002 &                        0.9389 $\pm$ 0.0008 \\
 SCC2016 &         0.893 $\pm$ 0.001 &             0.923 $\pm$ 0.001 &  0.8938 $\pm$ 0.0009 &    0.925 $\pm$ 0.001 &                               0.9211 $\pm$ 0.0007 &  \colorbox{lightgray}{\textbf{0.946}} &                      0.901 $\pm$ 0.001 &                          0.925 $\pm$ 0.001 \\
   ACMv9 &         0.926 $\pm$ 0.001 &            0.9350 $\pm$ 0.0007 &    0.919 $\pm$ 0.001 &  0.9352 $\pm$ 0.0001 &                               0.9254 $\pm$ 0.0006 &  \colorbox{lightgray}{\textbf{0.968}} &                      0.941 $\pm$ 0.001 &                        0.9525 $\pm$ 0.0007 \\
      EU &         0.795 $\pm$ 0.007 &           0.9297 $\pm$ 0.0004 &     0.760 $\pm$ 0.007 &  0.9264 $\pm$ 0.0008 &                               0.9169 $\pm$ 0.0006 &                        \textbf{0.944} &                      0.926 $\pm$ 0.007 &  \colorbox{lightgray}{0.9619 $\pm$ 0.0006} \\
     DNC &         0.766 $\pm$ 0.003 &    \textbf{0.929 $\pm$ 0.002} &     0.730 $\pm$ 0.001 &  0.8566 $\pm$ 0.0003 &                                 0.913 $\pm$ 0.001 &                                 0.919 &                       0.890 $\pm$ 0.006 &    \colorbox{lightgray}{0.939 $\pm$ 0.002} \\
    Wiki &           0.68 $\pm$ 0.02 &                0.70 $\pm$ 0.02 &      0.63 $\pm$ 0.01 &      0.63 $\pm$ 0.02 &    \colorbox{lightgray}{\textbf{0.83 $\pm$ 0.01}} &                                 0.801 &                        0.73 $\pm$ 0.01 &                            0.76 $\pm$ 0.02 \\
     UCS &         0.754 $\pm$ 0.005 &  \textbf{0.8762 $\pm$ 0.0008} &    0.717 $\pm$ 0.003 &  0.8558 $\pm$ 0.0008 &                                 0.844 $\pm$ 0.002 &                                  0.850 &                      0.904 $\pm$ 0.005 &   \colorbox{lightgray}{0.9530 $\pm$ 0.0008} \\
     Blg &         0.784 $\pm$ 0.001 &           0.9312 $\pm$ 0.0001 &    0.767 $\pm$ 0.002 &  0.9321 $\pm$ 0.0003 &                      \textbf{0.9334 $\pm$ 0.0001} &                                 0.924 &                      0.824 $\pm$ 0.001 &  \colorbox{lightgray}{0.9463 $\pm$ 0.0001} \\
     Ctl &           0.56 $\pm$ 0.03 &               0.66 $\pm$ 0.02 &      0.57 $\pm$ 0.03 &      0.67 $\pm$ 0.02 &     \colorbox{lightgray}{\textbf{0.70 $\pm$ 0.03}} &                                 0.574 &                        0.56 $\pm$ 0.03 &                            0.66 $\pm$ 0.02 \\
  FAA &         0.576 $\pm$ 0.003 &             0.589 $\pm$ 0.002 &    0.543 $\pm$ 0.007 &    0.535 $\pm$ 0.004 &                                 0.607 $\pm$ 0.003 &  \colorbox{lightgray}{\textbf{0.779}} &                      0.592 $\pm$ 0.002 &                          0.595 $\pm$ 0.002 \\
\midrule
Avg. & 0.749 $\pm$ 0.007 &\textbf{0.831 $\pm$ 0.004} &  0.700 $\pm$ 0.007&0.796 $\pm$ 0.006 & 0.813 $\pm$ 0.006 &0.823 &0.804 $\pm$ 0.005 &\colorbox{lightgray}{0.867 $\pm$ 0.006} \\
\bottomrule
\end{tabular}
}
\end{adjustbox}
\label{tab:apx_RD_auc}
\end{table*}

\clearpage
\newpage
% Bibliography
%\bibliographystyle{ScienceAdvances}
%\bibliographystyle{apsrev4-2}
\bibliography{bibliography}

%\end{widetext}
\end{document}